\documentclass[a4paper,11pt]{article}
\pdfoutput=1 

\usepackage{jheppub} 

\usepackage[T1]{fontenc}
\usepackage{multirow,subcaption}

\allowdisplaybreaks
 
\definecolor{nicered}{rgb}{0.7,0.1,0.1}
\definecolor{nicegreen}{rgb}{0.1,0.5,0.1}
\definecolor{niceblue}{rgb}{0.0,0.1,0.7}
\hypersetup{colorlinks,citecolor=niceblue,linkcolor=niceblue,urlcolor=niceblue}

\usepackage[normalem]{ulem}

\def \bm#1{\mbox{\boldmath$#1$\unboldmath}}
\def \beq{\begin{equation}}
\def \eeq{\end{equation}}
\def \bea{\begin{eqnarray}}
\def \eea{\end{eqnarray}}

\preprint{MPP-2022-65}

\title{\boldmath On Drell-Yan production of scalar leptoquarks coupling to heavy-quark flavours}

\author[a]{Ulrich Haisch,}
\author[a,\, b]{Luc Schnell}
\author[a,\, b]{and Stefan Schulte}

\affiliation[a]{Max Planck Institute for Physics, \\ F{\"o}hringer Ring 6, 80805 M{\"u}nchen, Germany}
\affiliation[b]{Technische Universit{\"a}t M{\"u}nchen, Physik-Department, \\ James-Franck-Strasse 1, 85748 Garching, Germany}

\emailAdd{haisch@mpp.mpg.de}
\emailAdd{schnell@mpp.mpg.de}
\emailAdd{sschulte@mpp.mpg.de}

\abstract{Given the hints of lepton-flavour non-universality in semi-leptonic~$B$ decays, leptoquark (LQ) models with sizeable couplings to heavy-quark flavours are enjoying a renaissance. While such models are subject to stringent constraints from low-energy experiments also bounds from non-resonant dilepton searches at the Large Hadron Collider~(LHC) turn out to be phenomenologically relevant. Based on the latest LHC dilepton analyses corresponding to an integrated luminosity of around~$140 \, {\rm fb}^{-1}$ of proton-proton collisions at~$\sqrt{s} = 13 \, {\rm TeV}$, we present improved limits on the scalar~LQ couplings that involve heavy-quark flavours and light or heavy~dileptons. In particular, we show that effects beyond the leading order that are related to real QCD emissions are relevant in this context, since the inclusion of additional heavy-flavoured jets notably improves the exclusion limits that derive from the high-mass dilepton tails. The impact of electroweak corrections and interference effects between signal and background is also analysed. Within the {\tt POWHEG-BOX}~framework we~provide a dedicated Monte Carlo code that allows for an on-the-fly signal event generation including all the~LQ corrections considered in this article. }

\begin{document} 
\maketitle
\flushbottom

\section{Introduction}
\label{sec:introduction}

Measurements of Drell-Yan~(DY) production represent a pillar of the research programme at the Large Hadron Collider~(LHC). These~searches are possible owing to the clean and well reconstructable experimental signature with excellent detection efficiency. While in the context of physics beyond the Standard Model~(SM) both~ATLAS and CMS have mostly investigated resonant dilepton signatures, recently also searches for non-resonant phenomena leading to dilepton final states~\cite{ATLAS:2020zms,ATLAS:2020yat,CMS:2021ctt,ATLAS:2021mla,CMS:2022uul,CMS-PAS-HIG-21-001} have been performed. 

Investigating non-resonant phenomena in DY~production involving both light ($e^+ e^-$ or~$\mu^+ \mu^-$) and heavy~($\tau^+ \tau^-$)~dilepton pairs is theoretically well motivated~\cite{Faroughy:2016osc,Raj:2016aky,Greljo:2017vvb,Allanach:2017bta,Dorsner:2018ynv,Afik:2018nlr,Bansal:2018eha,Allanach:2018odd,Schmaltz:2018nls,Mandal:2018kau,Baker:2019sli,Choudhury:2019ucz,Angelescu:2020uug,Bhaskar:2021pml,Crivellin:2021rbf,Cornella:2021sby,Crivellin:2021egp,Crivellin:2021bkd,Garland:2021ghw,Crivellin:2022mff,Azatov:2022itm} given the persisting hints of lepton-flavour universality violation that have been observed in the~$b \to c\ell \nu$~\cite{BaBar:2012obs,BaBar:2013mob,LHCb:2015gmp,LHCb:2017smo,LHCb:2017rln,Belle:2019gij} and~$b \to s \ell^+ \ell^-$~\cite{LHCb:2017avl,LHCb:2019hip,Belle:2019oag,BELLE:2019xld,LHCb:2021trn} systems. An apparent link between these flavour anomalies and non-resonant modifications in DY~dilepton distributions arises in leptoquark~(LQ) models with sizeable couplings to heavy-quark flavours, where~$t$-channel~LQ~exchange contributes to~$p p \to \ell^+ \ell^-$ production at the tree level. In~LQ models of this type the enhancement of new-physics effects at high energies can be utilised to curb the limited precision of the existing DY~dilepton measurements, allowing the bounds obtained in this way to be both complementary and competitive with those derived from precision low-energy data. It has also been noticed~\cite{ATLAS:2020zms,ATLAS:2021mla,CMS-PAS-HIG-21-001,Afik:2018nlr,Choudhury:2019ucz,Altmannshofer:2017poe,Iguro:2017ysu,Abdullah:2018ets,Marzocca:2020ueu,Endo:2021lhi} that the sensitivities to models that provide an explanation of the anomalies in~semileptonic $B$ decays may be improved by requiring an additional jet containing the decay of a~$B$ hadron~($b\hspace{0.4mm}$-jet) in the final state. 

The main goal of this article is to refine the theoretical description of DY~production in scalar~LQ models (see also~\cite{Kramer:1997hh,Kramer:2004df,Hammett:2015sea,Mandal:2015lca,Borschensky:2020hot,Buonocore:2020erb,Buonocore:2020nai,Greljo:2020tgv,Haisch:2020xjd,Borschensky:2021hbo} for publications similar in spirit). To this purpose we calculate the next-to-leading order~(NLO) QCD corrections to~$pp \to \ell^+ \ell^-$ production. This~computation involves the evaluation of the real and virtual corrections to the~$t$-channel Born-level contribution as well as the calculation of resonant single-LQ production followed by the decay of the~LQ. Such a calculation has been performed in the case of first- and second-generation~LQs already~in the article~\cite{Alves:2018krf} but not for third-generation~LQs, which is the main focus here. Besides QCD corrections we also consider the phenomenological impact of electroweak~(EW) corrections and study~the size of interference effects between the~leading~order~(LO)~LQ~signal and the LO SM background. These fixed-order predictions are consistently matched to a parton shower~(PS) employing the {\tt POWHEG} method~\cite{Nason:2004rx,Frixione:2007vw} as automatised in the {\tt POWHEG-BOX}~\cite{Alioli:2010xd}. This~allows for a realistic exclusive description of  DY dilepton~processes in scalar~LQ models at the level of hadronic events. In particular, our {\tt POWHEG} implementation can generate events with one additional parton from the matrix element calculation without the need to introduce a merging or matching scale. This~enables us to study~the constraints on scalar~LQ models that derive from the DY~searches in high-mass dimuon~($\mu^+ \mu^-$) final states without~\cite{ATLAS:2020yat} and with a~$b\hspace{0.4mm}$-jet~\cite{ATLAS:2021mla}. Finally, we also determine the restrictions that the latest ditau~($\tau^+ \tau^-$) search~\cite{CMS-PAS-HIG-21-001} put on scalar~LQ models studying two different~$b\hspace{0.4mm}$-jet categories. Based on our DY~analyses we are able to derive improved limits on the parameter space of third-generation scalar~LQ models using the full LHC~Run~II integrated luminosity of around~$140 \, {\rm fb}^{-1}$ obtained for proton-proton~($pp$) collisions at a centre-of-mass~energy of~$\sqrt{s} = 13 \, {\rm TeV}$.

The remainder of this article is organised in the following way. In Section~\ref{sec:theory} we specify the structure of the~LQ interactions that we consider in this work. Section~\ref{sec:calculation}~briefly describes the basic ingredients of the calculations of the different~LQ contributions to~DY~production and their implementation into the {\tt POWHEG-BOX}. The impact of the different types of~LQ corrections on the kinematic distributions in~$pp \to \ell^+ \ell^-$ production is presented in~Section~\ref{sec:analysis}. Our recasts of the LHC searches~\cite{ATLAS:2020yat,ATLAS:2021mla,CMS-PAS-HIG-21-001} are discussed in~Section~\ref{sec:limits}, where we also derive improved limits on the Yukawa couplings and masses of third-generation scalar~LQs. We~conclude and present an outlook in~Section~\ref{sec:conclusions}. Constraints on the parameter space of second-generation scalar LQs are provided in the supplementary material that can be found in~Appendix~\ref{app:second}.

\section{Theoretical framework}
\label{sec:theory}

LQs are hypothetical coloured bosons that carry both baryon and lepton number~\cite{Pati:1974yy}. They~therefore often emerge in beyond the SM~(BSM) models that unify matter~\cite{Pati:1973uk}. Since~any viable theory of unification has to reduce at low energies to the~SM such that the particle phenomenology observed in experiments is reproduced, scalar~LQs can only appear in five different representations~\cite{Buchmuller:1986zs,Dorsner:2016wpm}. In order to illustrate the possible effects of scalar~LQ contributions to DY~dilepton processes, we focus on the following simplified~LQ~model 
\beq \label{eq:SLQ}
{\cal L} \supset Y_{u \ell}\, \bar{u}^{\hspace{0.25mm} c} \ell \, S_1^{\dagger} + Y_{d \ell}\, \bar{d}^{\hspace{0.25mm} c} \ell \, \tilde{S}_1^{\dagger} + \text{h.c.} \,,
\eeq
where~$u, d$ and~$\ell$ represent the right-handed up-type, down-type quarks and charged lepton fields, respectively, and the superscript~$c$ denotes charge conjugation. The fermionic~SM fields are understood to be mass eigenstates,~i.e.~the states that lead to diagonal SM Yukawa coupling matrices after spontaneous EW symmetry breaking. The couplings~$Y_{u \ell}$ and~$Y_{d \ell}$ are complex~$3 \times 3$ matrices in flavour space, while the fields~$S_1$ and~$\tilde S_1$ correspond to the two~$SU(2)_L$~LQ singlets allowed by gauge invariance. Explicitly, the~LQ fields transform as~$S_1 \sim \left(3,1, -1/3 \right)$ and~$\tilde S_1 \sim \left(3,1, -4/3\right)$ under the full~$SU(3)_C \times SU(2)_L \times U(1)_Y$ SM~gauge~group. Notice that the size of the modifications in~$pp \to \ell^+ \ell^-$ production due to~LQ exchange depends primarily on the flavour structure and the magnitude of the couplings~$Y_{u \ell}$ and~$Y_{d \ell}$. However, once interference effects between the~LQ signal and the~SM background are considered also the representation of the~LQ plays a role because~the interference pattern depends on the quantum numbers of the exchanged~LQ~\cite{Crivellin:2021rbf}. In fact, in the case of~$S_1$~$\big ($$\tilde S_1$$\big )$ it turns out that the above Lagrangian gives rise to destructive~(constructive) interference of the~LQ signal with the SM~DY~background. The interactions~(\ref{eq:SLQ}) can therefore be used as a template to cover the full space of scalar~LQ models which entails besides the~$SU(2)_L$ singlets~$S_1$ and~$\tilde S_1$ the~$SU(2)_L$ doublets~$S_2$ and~$\tilde S_2$ and an~$SU(2)_L$ triplet~$S_3$. In~this context, we add that the fields~$S_2$ and~$S_3$ lead to constructive interference, while~$\tilde S_2$ interferes destructively with the SM~DY~background. 

\section{Calculation in a nutshell}
\label{sec:calculation}

\begin{figure}[t!]
\begin{center}
\includegraphics[width=0.75\textwidth]{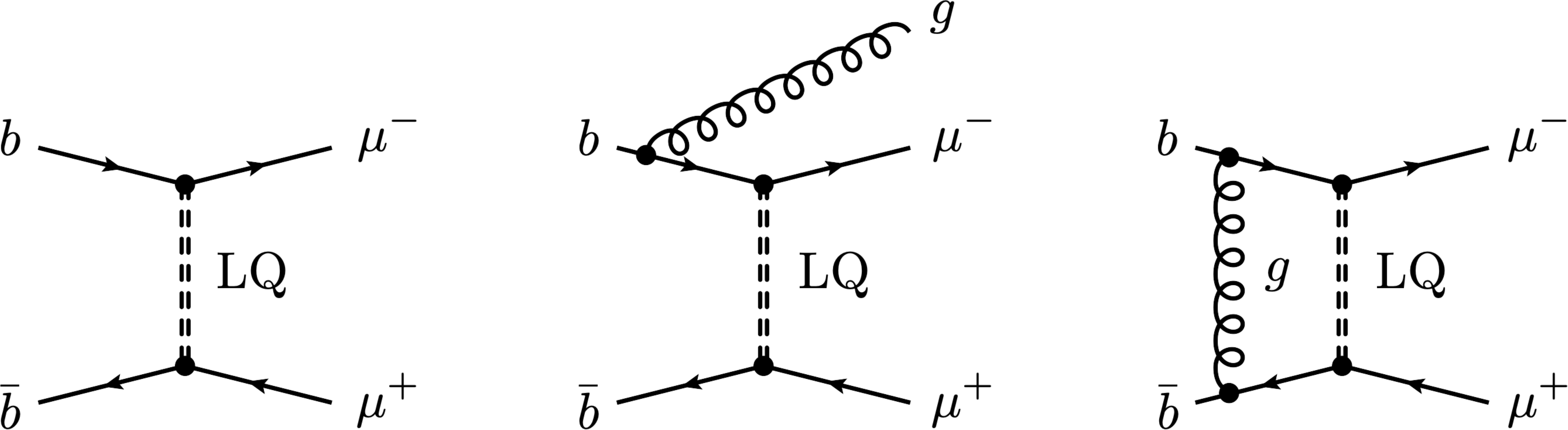}
\end{center}
\vspace{-1mm} 
\caption{\label{fig:diagrams1} Examples of~LQ contributions to DY~dimuon production initiated by bottom-quark fusion. The left Feynman diagram describes the tree-level process involving~$t$-channel~LQ exchange, while the middle (right) graph represents the corresponding real (virtual) QCD corrections. See~main text for further details.}
\end{figure}

\begin{figure}[t!]
\begin{center}
\includegraphics[width=0.9\textwidth]{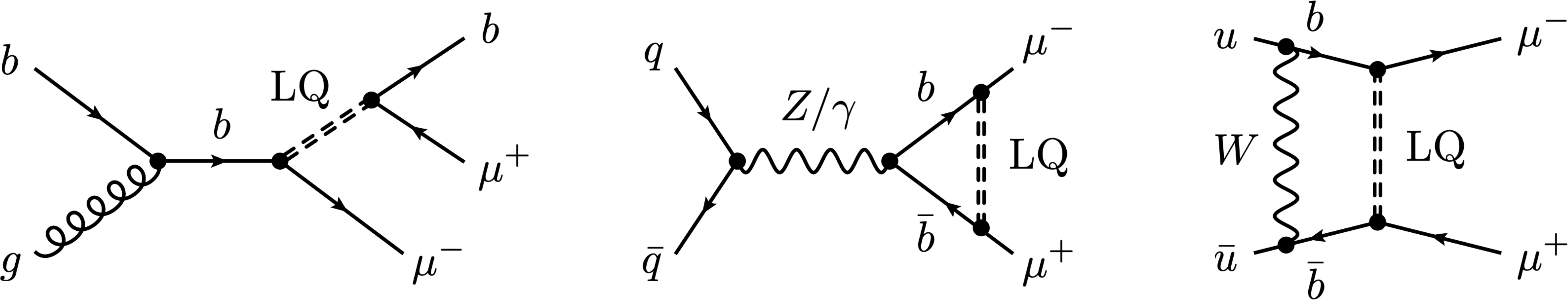}
\end{center}
\vspace{-2mm} 
\caption{\label{fig:diagrams2} An assortment of~LQ contributions to DY~dimuon production that arise beyond the~leading order in perturbation theory. The left graph is an example of resonant single-LQ production followed by the decay of the~LQ, whereas the middle and right diagram represent EW~corrections involving~LQ exchange. For additional explanations consult the main text.}
\end{figure}

Figures~\ref{fig:diagrams1} and~\ref{fig:diagrams2} display representative Feynman diagrams inducing~DY~dimuon production in the presence of~(\ref{eq:SLQ}). The first figure shows the tree-level contribution involving~$t$-channel~LQ exchange (left) and the corresponding real (middle) and virtual (right) QCD~corrections. Notice that all depicted contributions are initiated by bottom-quark~($b \bar b$) fusion\footnote{Throughout this article we work in the five-flavour scheme, where charm- and bottom-quarks are considered as partons in the proton and as such have a corresponding parton distribution function~(PDF).} and that the exchanged~LQ is an~$\tilde S_1$. An assortment of~LQ contributions to DY~dimuon production that arise beyond the~LO in perturbation theory is given in the second figure. The~left Feynman diagram gives rise to resonant single-LQ production with subsequent decay of the~LQ to a pair of a bottom quark and an anti-muon,~i.e.~$gb \to \tilde S_1 \hspace{0.25mm} \mu^-$ with~$ \tilde S_1 \to b \hspace{0.25mm} \mu^+$. Notice that graphs of this type as well as the real and virtual corrections shown in~Figure~\ref{fig:diagrams1} all represent a~${\cal O} (\alpha_s)$ correction to the inclusive DY~dilepton production rate. In order to achieve NLO accuracy in QCD one therefore has to include all three classes of graphs. Notice that the diagrams in~Figure~\ref{fig:diagrams1} and the left graph in~Figure~\ref{fig:diagrams2} with bottom replaced by charm quarks arise in the case of the~LQ singlet~$S_1$.

Besides QCD corrections to~$pp \to \ell^+ \ell^-$ we also consider EW effects to DY~production in our article. Two prototype graphs of this kind are shown in the centre and on the right-hand side of Figure~\ref{fig:diagrams2}. The first type of diagrams encodes the virtual corrections to the~$Z\ell^+\ell^-$ and~$\gamma \ell^+\ell^-$ vertices involving the exchange of an~LQ. These vertex corrections appear both in the initial and the final state. The second type of EW corrections is associated to one-loop Feynman graphs with~$W$-boson exchange. Notice that due to the structure of~(\ref{eq:SLQ}), which only involves right-handed fermionic fields, EW contributions of the latter kind are strongly chirally suppressed by small SM Yukawa couplings. In the case of DY~production by heavy-quark fusion these corrections furthermore involve small Cabibbo-Kobayashi-Maskawa matrix elements. We therefore do not include EW corrections related to~$W$-boson exchange in our analysis. Likewise, we also do not consider EW contributions due to SM~Higgs-boson exchange, because these corrections are again insignificant as they are proportional to small SM Yukawa couplings. 

The third kind of quantum effects that we consider in our work is the interference between the~LQ and the SM contributions to tree-level $q \bar q \to \ell^+ \ell^-$ scattering. We treat these contributions at the LO in perturbation theory, which means that our {\tt POWHEG-BOX} implementation contains the squared matrix elements built from the $t$-channel~LQ contribution and the SM corrections involving $Z$-boson or photon exchange in the $s$-channel. 

All matrix elements are computed using conventional dimensional regularisation for both ultraviolet~(UV) and infrared~(IR) singularities. The actual generation and computation of squared matrix elements relies on the {\tt Mathematica} packages {\tt FeynRules}~\cite{Alloul:2013bka}, {\tt FeynArts}~\cite{Hahn:2000kx}, {\tt FormCalc}~\cite{Hahn:2016ebn}, {\tt LoopTools}~\cite{Hahn:1998yk} and {\tt Package-X}~\cite{Patel:2015tea}.  Our calculation of NLO~QCD and EW effects is performed  in the on-shell scheme. In order to deal with the soft and collinear singularities of the real corrections to the~$t$-channel~LQ~exchange contribution, cf.~the~middle diagram in~Figure~\ref{fig:diagrams1}, and to cancel the IR~poles of the one-loop virtual corrections, cf.~the~right diagram in~Figure~\ref{fig:diagrams1}, we~exploit the general implementation of the Frixione-Kunszt-Signer subtraction~\cite{Frixione:1995ms,Frixione:1997np} within the~{\tt POWHEG-BOX} framework. For this purpose, the full {\tt POWHEG-BOX} machinery is used that automatically builds the soft and collinear counterterms and remnants, and also checks the behaviour in the soft and collinear limits of the real squared matrix elements against their soft and collinear approximations. Notice that the real NLO~QCD contributions that describe resonant single-LQ production with subsequent decay of the~LQ are IR~finite and therefore do not require a subtraction~(cf.~the~left diagram in~Figure~\ref{fig:diagrams2}). Our Monte~Carlo~(MC) code therefore allows to achieve NLO+PS accuracy for DY~dilepton production in any scalar~LQ model described by~(\ref{eq:SLQ}). In particular, our {\tt POWHEG} implementation is able to generate events with one additional parton from the matrix element calculation without the need to introduce a merging or matching scale. Two-jet events are instead exclusively generated by the PS in our MC setup. 

Let us~finally add that the results of our calculation of the virtual corrections to the~$Z\ell^+\ell^-$ and~$\gamma \ell^+\ell^-$ vertices involving the exchange of a~LQ can be shown to resemble the leading terms in the heavy-mass expansion of the corresponding form factors given in the publication~\cite{Crivellin:2020mjs}. This comparison serves as a useful cross-check of our computation.

\section{Phenomenological analyses}
\label{sec:analysis}

In this section we discuss the numerical impact of the different types of~LQ corrections on the kinematic distributions that are most relevant for the existing LHC searches for non-resonant BSM physics in dilepton final states. The case of light and heavy dilepton pairs is discussed separately and in both cases signatures with~no or one $b\hspace{0.4mm}$-jet are considered. All~results shown in the following are obtained assuming~$pp$ collisions at $\sqrt{s} =13 \, {\rm TeV}$, they employ {\tt NNPDF40\_nlo\_as\_01180} PDFs~\cite{Ball:2021leu} and {\tt Pythia~8}~\cite{Sjostrand:2014zea} is used to shower the events. Effects~from hadronisation, underlying event modelling or QED effects in the PS are not included in our MC simulations. 

\subsection{Inclusive light dilepton final states}
\label{sec:inclusivelightpheno}

\begin{figure}[t!]
\begin{center} 
\includegraphics[width=0.475\textwidth]{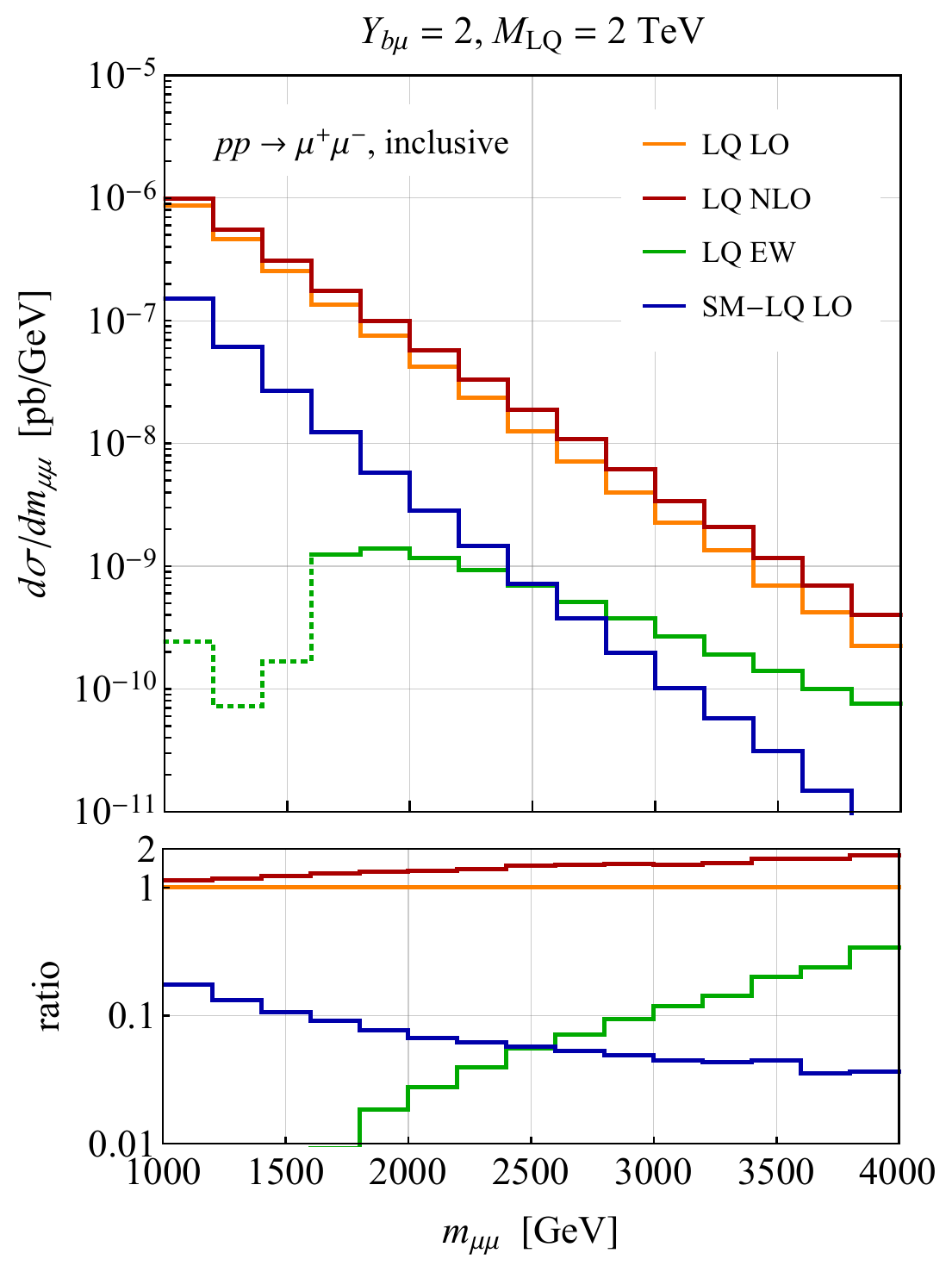} \quad 
\includegraphics[width=0.475\textwidth]{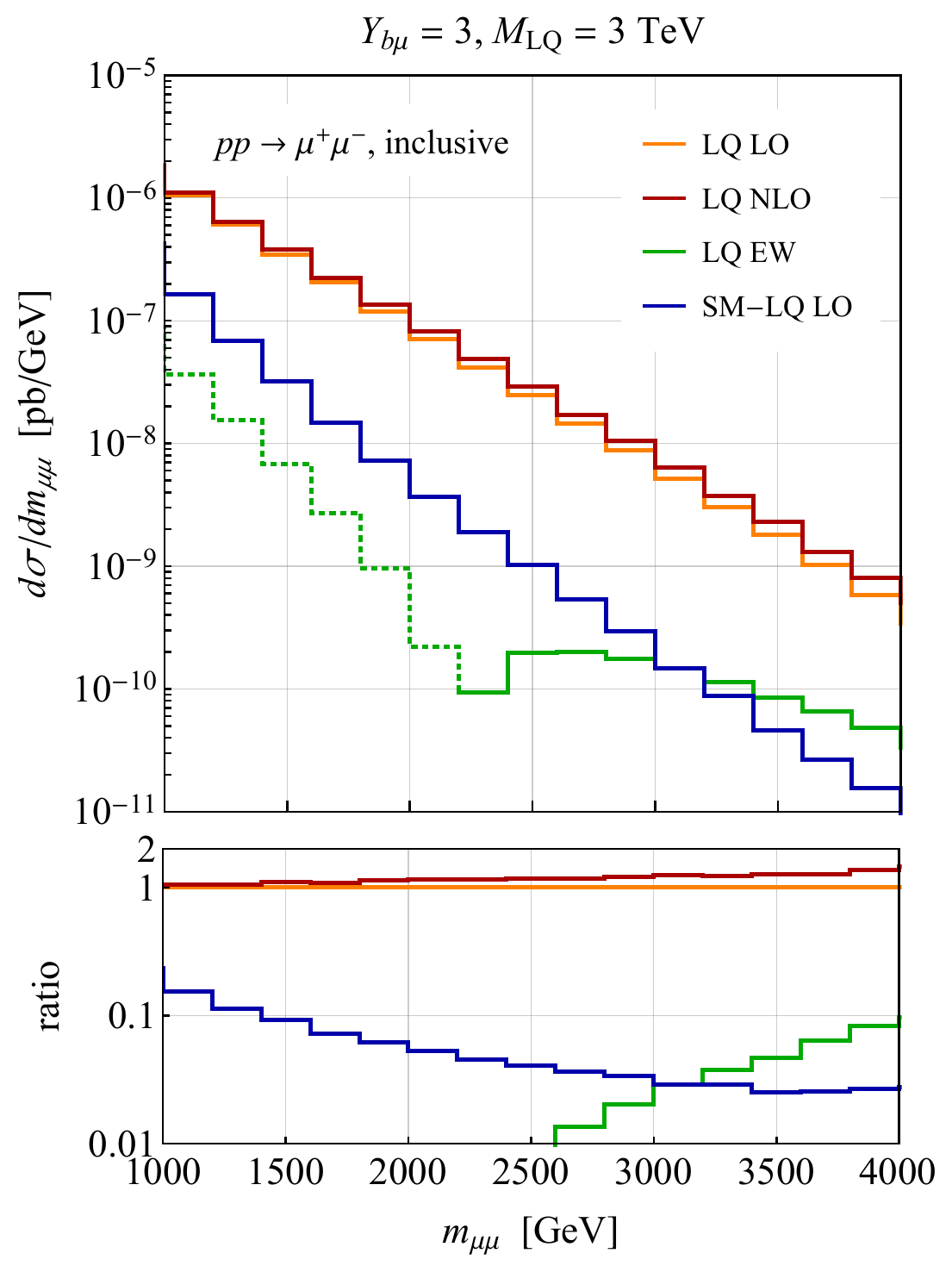}
\end{center}
\vspace{-4mm} 
\caption{\label{fig:pheno1} Inclusive $pp \to \mu^+ \mu^-$ production cross sections as a function of $m_{\mu \mu}$. The~left~(right) plot shows the results for $Y_{b\mu} = 2, M_{\rm LQ} = 2 \, {\rm TeV}$ $\big($$Y_{b\mu} = 3, M_{\rm LQ} = 3 \, {\rm TeV}$$\big)$. The~LQ couplings not specified in the headline of the plots are set to zero. The yellow and red curves correspond to the~LQ distributions at the LO~(LQ~LO) and the~NLO~(LQ~NLO) in QCD, respectively, while the green and blue histograms illustrate the impact of~EW~effects~(LQ~EW) and the size of the interference effects between the~LQ~signal and the SM background~(SM-LQ~LO). The parts of the green curves that are dotted correspond to negative EW contributions to the differential cross sections. The~lower panels depict the ratios between the different~LQ contributions and the relevant~LQ~LO~distribution.}
\end{figure}

The simplest LHC searches for non-resonant DY~phenomena (see for instance~\cite{ATLAS:2020yat,CMS:2021ctt}) use inclusive measurements of the high-mass dielectron or dimuon~($m_{\mu \mu}$) tail to set constraints on non-SM physics. In~Figure~\ref{fig:pheno1} we present our results for the~LQ corrections to the~$m_{\mu \mu}$ spectrum in inclusive $pp \to \mu^+ \mu^-$ production adopting two benchmark choices for~$Y_{b\mu}$ and~$M_{\rm LQ}$. All other~LQ couplings are set to zero to obtain the results shown in~the~figure. The yellow and red curves in both plots correspond to the~LQ distributions at the LO~(LQ~LO) and the~NLO~(LQ~NLO) in QCD, respectively, while the green and blue histograms illustrate the impact of~EW~corrections~(LQ~EW) and the size of the interference effects between the~LQ~signal and the SM background~(SM-LQ~LO). From~the lower panel of the~left plot it is evident that for the choice $Y_{b\mu} = 2, M_{\rm LQ} = 2 \, {\rm TeV}$ the NLO~QCD effects play an important role in obtaining precise predictions as they amount compared to the tree-level~LQ predictions to around~50\%~(80\%) at $m_{\mu \mu} = 3 \, {\rm TeV}$ ($m_{\mu \mu} = 4 \, {\rm TeV}$). The~corresponding numbers in the case of $Y_{b\mu} = 3, M_{\rm LQ} = 3 \, {\rm TeV}$ are 25\% and 40\%. Higher-order~EW corrections are far less important than the NLO~QCD contributions at low invariant masses\footnote{Below the~LQ threshold the EW effects lead to a reduction of the differential DY~cross section. This~is~indicated in Figure~\ref{fig:pheno1} by the dotted green parts of the histograms.} but become relevant at high energies where they can lead to enhancements of the production rates of more than~30\% for $Y_{b\mu} = 2, M_{\rm LQ} = 2 \, {\rm TeV}$. This~feature is well-known~(cf.~for example~\cite{Ciafaloni:1998xg}) and due to the appearance of Sudakov logarithms of the form $\ln^2 \big ( m_{\mu \mu}^2/M_{\rm LQ}^2 \big )$ which are associated to virtualites $q^2 \simeq m_{\mu \mu}^2$ that are much larger than the mass of the~LQ entering the loop diagrams. The~double-logarithmic behaviour also explains why for $Y_{b\mu} = 3, M_{\rm LQ} = 3 \, {\rm TeV}$ the EW~corrections are less pronounced than in the case of $Y_{b\mu} = 2, M_{\rm LQ} = 2 \, {\rm TeV}$. Interference effects between the~LQ signal and the SM background amount in both cases to approximately 5\% in the high-mass tail of the $m_{\mu \mu}$ spectrum and are therefore only of minor importance. 

\begin{figure}[t!]
\begin{center} 
\includegraphics[width=0.475\textwidth]{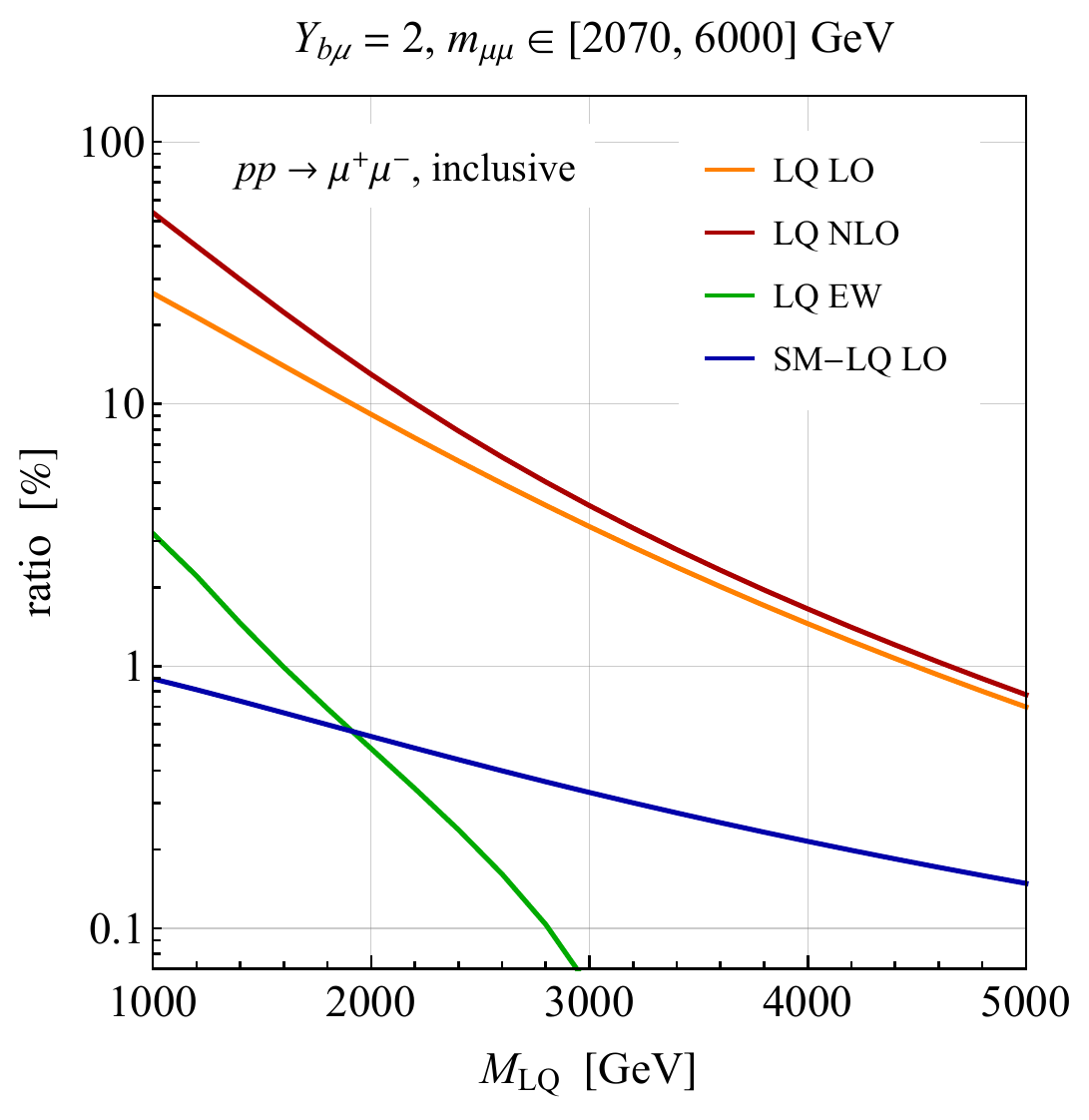} \quad 
\includegraphics[width=0.475\textwidth]{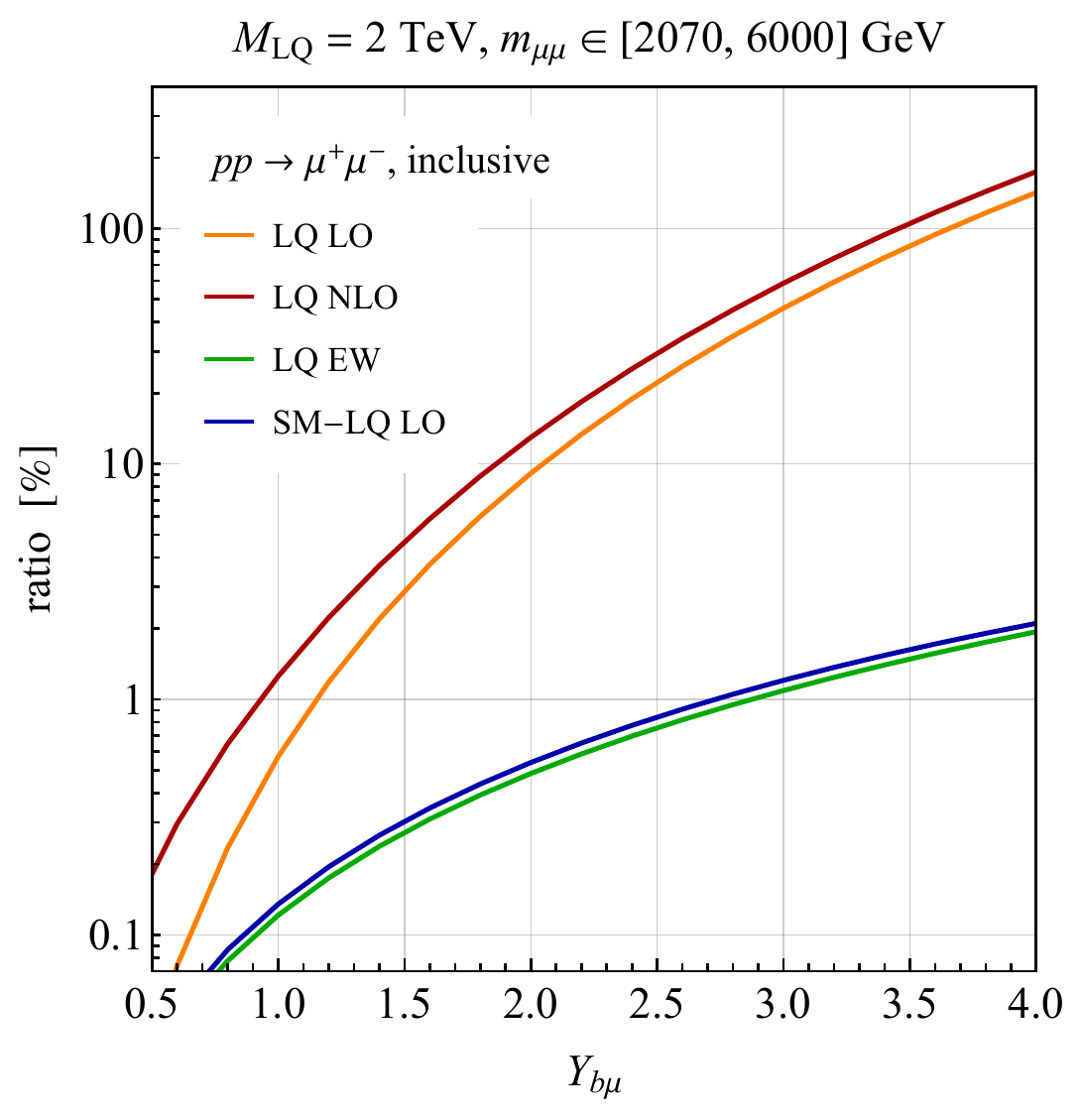}
\end{center}
\vspace{-4mm} 
\caption{\label{fig:pheno2} Ratios between the individual~LQ corrections and the~inclusive DY~SM background calculated at the NLO in QCD. The shown results correspond to the fiducial region defined by $p_{T, \hspace{0.25mm} \mu} > 30 \, {\rm GeV}$, $|\eta_\mu| < 2.5$ and $m_{\mu \mu} \in [2070, 6000] \, {\rm GeV}$. The~left (right) plot depicts the results as a function of $M_{\rm LQ}$ ($Y_{b\mu}$) for fixed $Y_{b\mu} =2$ ($M_{\rm LQ} = 2 \, {\rm TeV}$). The~colour coding and meaning of the different curves resembles those in Figure~\ref{fig:pheno1}. Additional details can be found in the main text.} 
\end{figure}

The results shown in Figure~\ref{fig:pheno1} already~suggest that in existing LHC searches for non-resonant phenomena in DY~distributions the phenomenological impact of EW and interference effects involving~LQs is limited. To further illustrate this point we display in~Figure~\ref{fig:pheno2} the ratios between the individual~LQ contributions and the inclusive DY~SM background in the fiducial region. The~normalisation is calculated at the NLO~in~QCD and we select events that contain two opposite-sign same-flavour~(OSSF) muons that are both required to have a transverse momentum of $p_{T, \hspace{0.25mm} \mu} > 30 \, {\rm GeV}$ and a pseudorapidity of $|\eta_\mu| < 2.5$ and their invariant mass must fall into the range $m_{\mu \mu} \in [2070, 6000] \, {\rm GeV}$. Detector efficiency corrections are not taken into account. Notice that this invariant mass window corresponds to the signal region~(SR) used by ATLAS in their recent non-resonant dimuon analysis~\cite{ATLAS:2020yat} assuming constructive signal-background interference. The~left panel displays our results as a function of $M_{\rm LQ}$ for fixed $Y_{b\mu} =2$. From this plot one sees that the relative size of the NLO~QCD corrections decreases for increasing~LQ mass. Numerically, we find relative effects of around 100\%, 20\% and 10\% at $M_{\rm LQ} = 1 \, {\rm TeV}$, $M_{\rm LQ} = 3 \, {\rm TeV}$ and $M_{\rm LQ} = 5 \, {\rm TeV}$. This feature is readily understood by noting that the NLO~QCD corrections related to $s$-channel single-LQ production followed by the decay of the~LQ,~cf.~the~left Feynman diagram in Figure~\ref{fig:diagrams2}, decouple faster than the real and virtual corrections to the $t$-channel Born-level~LQ contribution,~cf.~the middle and right graph in Figure~\ref{fig:diagrams1}. Another~property that is visible in the~left panel is the strong suppression of the EW corrections for increasing~$M_{\rm LQ}$. This is related to the fact that for heavy~LQs the enhancement of EW effects due to Sudakov double-logarithms is not at work in the considered SR. One~furthermore observes that both the EW and the interference effects represent only subleading corrections in the mass window $m_{\mu \mu} \in [2070, 6000] \, {\rm GeV}$, amounting to at most~3\% and below~1\%, respectively, in the shown~LQ mass range.  

The right panel in Figure~\ref{fig:pheno2} depicts our ratio predictions as a function of~$Y_{b\mu}$ setting the mass of the~LQ to $M_{\rm LQ} = 2 \, {\rm TeV}$. We see that the relative size of the NLO~QCD corrections decreases for increasing coupling strength. Compared to the tree-level~LQ result the higher-order QCD effects amount to around 440\%, 40\% and 20\% for $Y_{b\mu}=0.5$, $Y_{b\mu}=2$ and $Y_{b\mu}=4$. This behaviour can be understood by realising that the~squared amplitude of the $t$-channel Born-level contribution scales as $|Y_{b \mu}|^4$, while the resonant single-LQ production rate is proportional to $|Y_{b \mu}|^2$. One notices furthermore that the relative~LQ~EW and SM-LQ~LO modifications both depend quadratically on $|Y_{b \mu}|$. One again sees that both the~EW as well as the interference contributions are numerically subleading even for large couplings~$Y_{b \mu}$ where they just reach the level of~1\%. 

\subsection[Light dilepton final states with one $b\hspace{0.4mm}$-jet]{Light dilepton final states with one $\bm{b}\hspace{0.4mm}$-jet}
\label{sec:bjetlightpheno}

\begin{figure}[t!]
\begin{center} 
\includegraphics[width=0.475\textwidth]{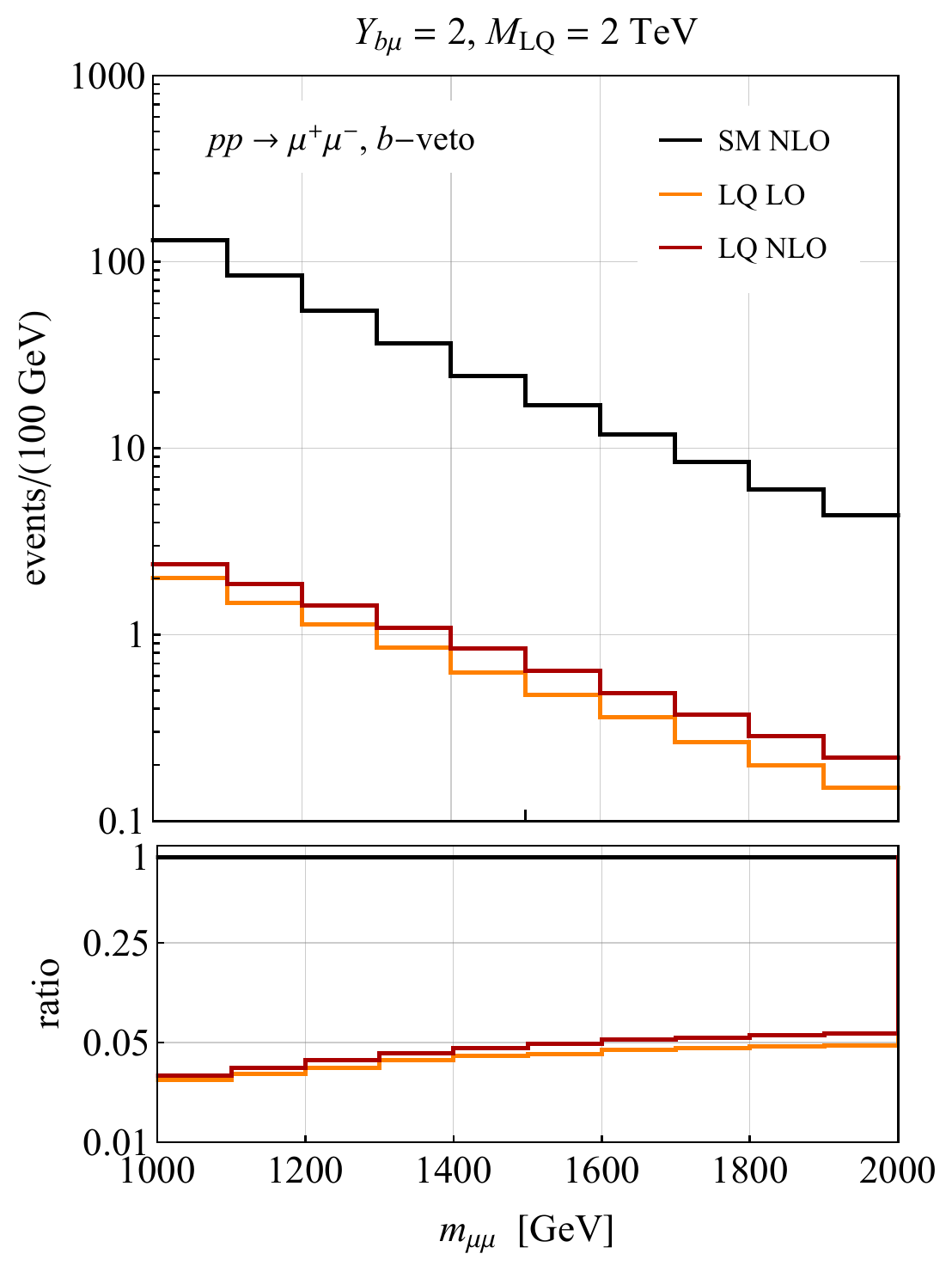} \quad 
\includegraphics[width=0.475\textwidth]{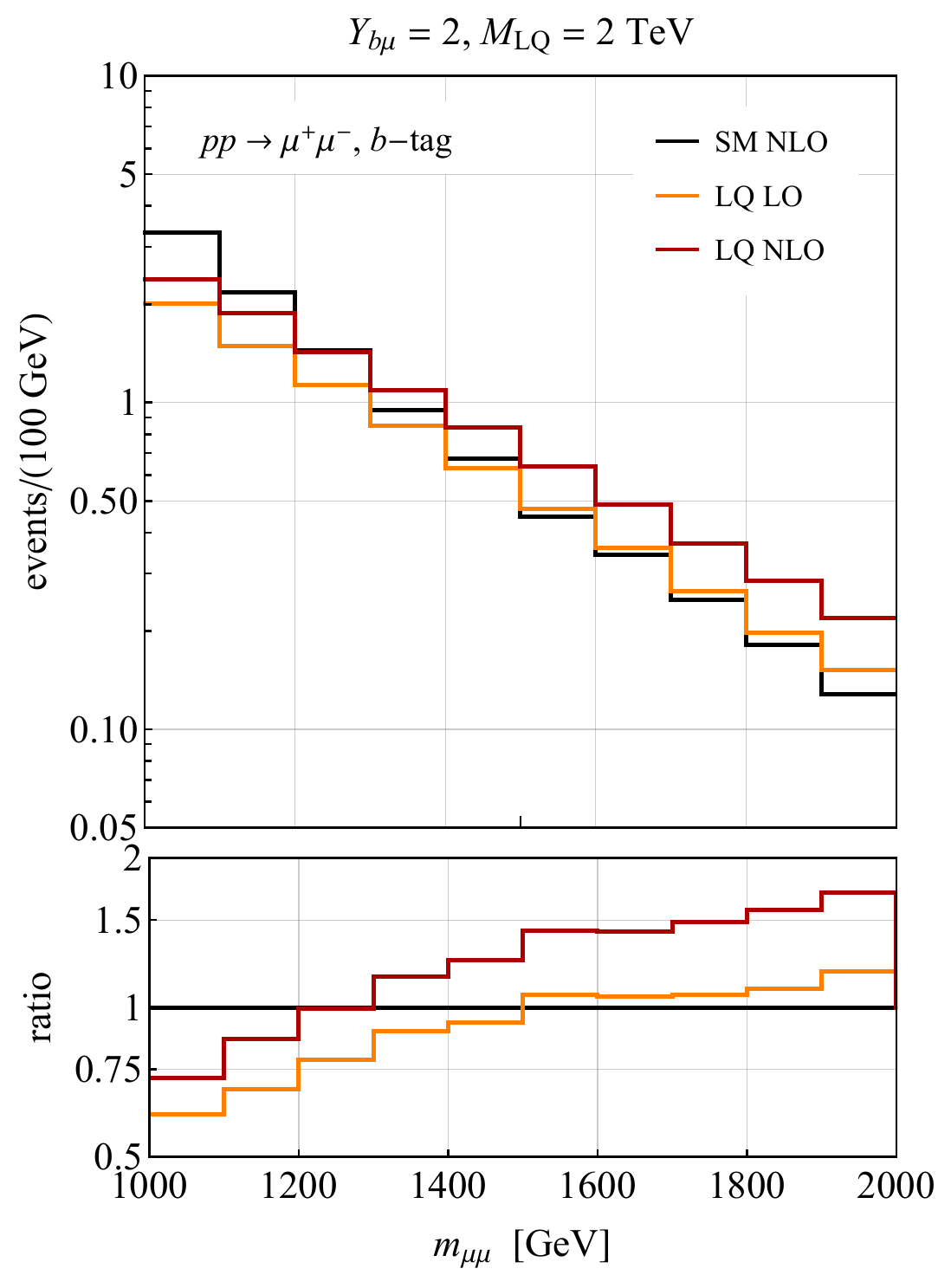}
\end{center}
\vspace{-4mm} 
\caption{\label{fig:pheno3} Inclusive $pp \to \mu^+ \mu^-$ production cross sections as a function of $m_{\mu \mu}$ for two different~$b\hspace{0.4mm}$-jet categories. The~left~(right) plot shows the results for $Y_{b\mu} = 2, M_{\rm LQ} = 2 \, {\rm TeV}$ imposing a $b\hspace{0.4mm}$-veto~($b\hspace{0.4mm}$-tag). The black, yellow and red curves correspond to the~SM results obtained at NLO in QCD~(SM~NLO), the~LQ LO and the~LQ~NLO predictions, respectively. All~results assume an integrated luminosity of $139 \, {\rm fb}^{-1}$. The~lower panels display the ratios between the different~LQ contributions and the corresponding SM~NLO spectrum. Further~details can be found in the main~text.}
\end{figure}

Inspired by the $b \to s \ell^+ \ell^-$ anomalies also LHC~searches for final states with two OSSF leptons and exactly one $b\hspace{0.4mm}$-jet have been proposed~\cite{Afik:2018nlr} and recently performed by~ATLAS~\cite{ATLAS:2021mla}. In order to illustrate the improvement in sensitivity that is gained by targeting dilepton final states with additional $b\hspace{0.4mm}$-jets, we show in Figure~\ref{fig:pheno3} inclusive $pp \to \mu^+ \mu^-$ cross sections as a function of $m_{\mu \mu}$ employing two different~$b\hspace{0.4mm}$-jet categories. We~adopt the~LQ parameter choices~$Y_{b\mu} = 2, M_{\rm LQ} = 2 \, {\rm TeV}$ and consider $139 \, {\rm fb}^{-1}$ of integrated luminosity under LHC~Run~II conditions. Following the study~\cite{ATLAS:2021mla} events are selected with two OSSF muons that are both required to satisfy $p_{T, \hspace{0.25mm} \mu} > 30 \, {\rm GeV}$ and $|\eta_\mu| < 2.5$. Jets are reconstructed using the anti-$k_t$ algorithm~\cite{Cacciari:2008gp} with radius parameter $R=0.4$, as implemented in {\tt FastJet}~\cite{Cacciari:2011ma}, and need to fulfil $p_{T, \hspace{0.25mm} j} > 30 \, {\rm GeV}$ and $|\eta_j| < 2.5$. Jets originating from the hadronisation of a bottom or anti-bottom quark are identified~(i.e.~$b\hspace{0.4mm}$-tagged) adopting the performance of the ATLAS $b\hspace{0.4mm}$-tagging algorithm described in~\cite{ATLAS:2019bwq}. For the analyses performed in this subsection, a $b\hspace{0.4mm}$-tagging working point is chosen that yields a~$b\hspace{0.4mm}$-tagging efficiency of~77\% and a light-flavour jet rejection of~110. Detector effects are simulated by applying reconstruction and identification efficiency factors tuned to mimic the performance of the ATLAS detector. In~particular, muon candidates must fulfil the ATLAS quality selection criteria optimised for high-$p_T$ performance~\cite{ATLAS:2016lqx,ATLAS:2020gty}. The corresponding reconstruction and identification efficiency amount to around 75\% in the phase-space region of interest. Our~analysis is implemented into~{\tt MadAnalysis~5}~\cite{Conte:2012fm} and employs {\tt Delphes~3}~\cite{deFavereau:2013fsa} as a fast detector simulator. Applying our MC chain to the SM~NLO prediction obtained with the {\tt POWHEG-BOX}, we are able reproduce the SM~DY~background postfit $m_{\mu \mu}$ distribution in the SR provided by ATLAS~in~\cite{ATLAS:2021mla} at the level of 10\%. This`comparison represents as a non-trivial cross-check of our analysis.

In the left panel of Figure~\ref{fig:pheno3} our results for the inclusive DY~dimuon cross section with~no~$b\hspace{0.4mm}$-tagged jet~($b\hspace{0.4mm}$-veto) are presented. The black, yellow and red histogram display the~SM results obtained at~NLO in QCD~(SM~NLO), the~LQ LO and the~LQ~NLO predictions, respectively. The size of EW and interference effects is not shown in the figure because these corrections are both very small. One observes that in the $b\hspace{0.4mm}$-veto category the~LQ contributions to the differential rate amount to corrections of a few percent only, and that~NLO~QCD corrections modifying the~LO~LQ spectrum by around 10\% to 20\% in the shown $m_{\mu \mu}$ range. Requiring one $b\hspace{0.4mm}$-jet~($b\hspace{0.4mm}$-tag) in addition to the two OSSF muons changes the picture radically. This~is~illustrated on the right-hand side in~Figure~\ref{fig:pheno3}. In fact, the requirement of an additional $b\hspace{0.4mm}$-jet reduces the SM~background by roughly a factor of 35 largely independent of $m_{\mu \mu}$, while the $b\hspace{0.4mm}$-jet requirement has an effect of around $-60\%$ ($-15\%$) on the~signal strength in the considered~LQ realisation at $m_{\mu \mu} = 1\, {\rm TeV}$~($m_{\mu \mu} = 2\, {\rm TeV}$).  It is also visible that~the size of the~NLO~QCD corrections to the~LQ signal is larger in the case of the $b\hspace{0.4mm}$-tag than the $b\hspace{0.4mm}$-veto category, exceeding 25\% above approximately $m_{\mu \mu} = 1.5 \, {\rm TeV}$. This feature is explained by noting that~NLO~QCD contributions of the form $gb \to \tilde S_1 \hspace{0.25mm} \mu^-$ with~$ \tilde S_1 \to b \hspace{0.25mm} \mu^+$,~cf.~the~left diagram in~Figure~\ref{fig:diagrams2}, will mostly contribute to the $b\hspace{0.4mm}$-tag category. Similar statements apply to channels like $g b \to \mu^+ \mu^- b$ where the anti-bottom quark that partakes in the $t$-channel~LQ process $b \bar b \to \mu^+ \mu^-$ arises from splitting of an initial-state gluon. Notice however that while the latter type of corrections can be partly captured by a PS when applied to the~LO matrix elements, this is not the case for the former contribution associated to resonant single-LQ production In order to achieve an accurate exclusive description of  DY~dilepton~processes in~LQ models involving heavy-flavoured jets,~NLO+PS predictions as provided in our work are therefore called for. 

\subsection[Heavy dilepton final states with and without a $b\hspace{0.4mm}$-jet]{Heavy dilepton final states with and without a $\bm{b}\hspace{0.4mm}$-jet}
\label{sec:heavypheno}

Searches for signatures involving tau pairs in the final state such as those performed at LHC~Run~II~\cite{ATLAS:2020zms,CMS-PAS-HIG-21-001} are known~\cite{Faroughy:2016osc,Dorsner:2018ynv,Schmaltz:2018nls,Mandal:2018kau,Baker:2019sli,Choudhury:2019ucz,Cornella:2021sby} to provide strong constraints on~LQ models that address the $b \to c \ell \nu$ anomalies. In the following we will consider the recent~CMS~search~\cite{CMS-PAS-HIG-21-001} for $\tau^+ \tau^-$ final states with both taus decaying to hadrons $\big($$\tau^\pm_{\rm h}$$\big)$ as an example to illustrate the role that additional $b\hspace{0.4mm}$-jets play in analyses of this kind. To distinguish hadronic $\tau$ candidates from jets originating from the hadronisation of quarks and gluons, and from electrons or muons the $\tau$-tagger described in~\cite{CMS:2022prd} is employed. The used working points have an efficiency of approximately 50\%, 70\% and 70\% for identification in the case of jets, electrons and muons, respectively. The corresponding rejection factors are around 230, 20, and 770. Both hadronic $\tau$ candidates are required to have $p_{T, \tau} > 40 \, {\rm GeV}$ and $|\eta_{\tau}| < 2.1$, and the angular distance between them must be greater than $\Delta R_{\tau \tau} = 0.3$ in the pseudorapidity-azimuth space. Jets are clustered using the anti-$k_t$ algorithm with radius $R=0.4$. Jets with $p_{T,j} > 30 \, {\rm GeV}$ and $|\eta_j| < 4.7$ and $b\hspace{0.4mm}$-jets with $p_{T,b} > 20 \, {\rm GeV}$ and $|\eta_b| < 2.5$ are selected. To identify $b\hspace{0.4mm}$-jets we employ the CMS $b\hspace{0.4mm}$-tagging efficiencies stated in~\cite{CMS:2017wtu,Bols:2020bkb}. The used $b\hspace{0.4mm}$-tagging working point yields a~$b\hspace{0.4mm}$-tagging efficiency of around 80\% and a light-flavour jet rejection in the ballpark of 100. {\tt MadAnalysis~5} in combination with {\tt Delphes~3}~is again used to analyse the events and to simulate the detector effects. We have verified that applying our analysis to the SM~NLO DY~prediction, we are able reproduce the SM~DY~background as given~in~\cite{CMS-PAS-HIG-21-001} to within around 30\%.

To discriminate between signal and background, we consider the distributions of the total transverse mass defined as~\cite{ATLAS:2014vhc}
\begin{equation} \label{eq:mTtot}
m_T^{\rm tot} = \sqrt{m_T^2 (\vec{p}_T^{\; \tau_1}, \vec{p}_T^{\; \tau_2}) + m_T^2 (\vec{p}_T^{\; \tau_1}, \vec{p}_T^{\; \rm miss}) + m_T^2 (\vec{p}_T^{\; \tau_2}, \vec{p}_T^{\; \rm miss}) } \,,
\end{equation}
where $\tau_1$ ($\tau_2$) refers to the first (second) hadronic $\tau$ candidate and $\vec{p}_T^{\; \tau_1}$, $\vec{p}_T^{\; \tau_2}$ and $ \vec{p}_T^{\; \rm miss}$ are the vectors with magnitude $p_{T, \tau_1}$, $p_{T, \tau_2}$ and $E_{T, \rm miss}$. Here $E_{T, \rm miss}$ denotes the missing transverse energy constructed from the transverse momenta of all the neutrinos in the event. The transverse mass of two transverse momenta $p_{T,i}$ and $p_{T, j}$ entering~(\ref{eq:mTtot}) is given by 
\begin{equation} \label{eq:mT}
m_T (\vec{p}_T^{\; i}, \vec{p}_T^{\; j} ) = \sqrt{2 \hspace{0.25mm} p_{T, i} \hspace{0.5mm} p_{T, j} \left ( 1 - \cos \Delta \phi \right )} \,, 
\end{equation}
where $\Delta \phi$ is the azimuthal angular difference between the vectors $\vec{p}_T^{\; i}$ and $ \vec{p}_T^{\; j}$. 

\begin{figure}[t!]
\begin{center} 
\includegraphics[width=0.475\textwidth]{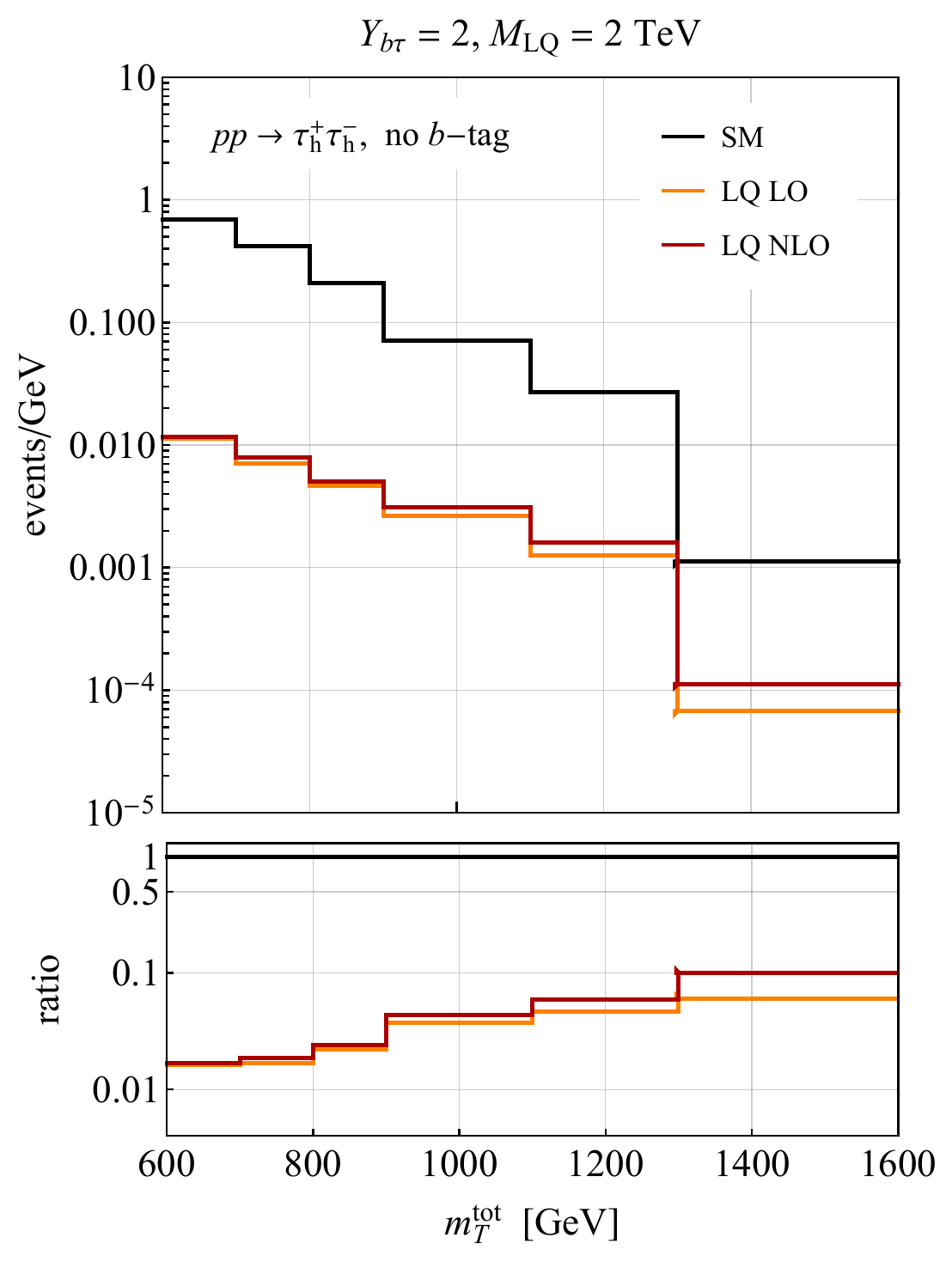} \quad 
\includegraphics[width=0.475\textwidth]{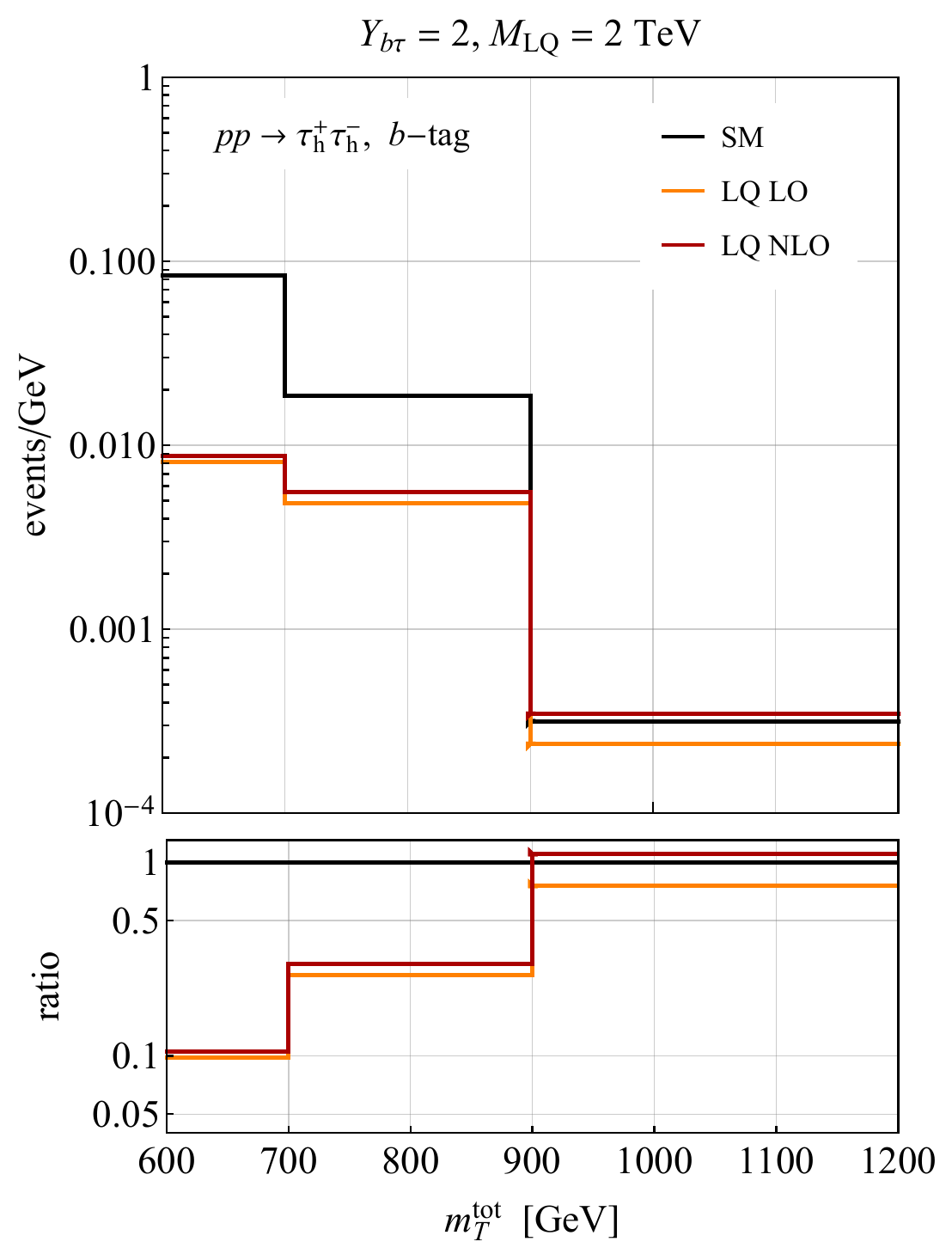}
\end{center}
\vspace{-4mm} 
\caption{\label{fig:pheno4} Distributions of $m_T^{\rm tot}$ in the no~$b\hspace{0.4mm}$-tag~(left~panel) and the~$b\hspace{0.4mm}$-tag (right~panel) categories in the $\tau_{\rm h}^+ \tau_{\rm h}^-$ final state. The~black curves correspond to the SM expectations of the DY~background provided by~CMS in the search~\cite{CMS-PAS-HIG-21-001} which is based on $138 \, {\rm fb}^{-1}$ of LHC~Run~II data The yellow and red curves instead represent the~LQ~LO and~LQ~NLO results assuming $Y_{b \tau} = 2, M_{\rm LQ} = 2 \, {\rm TeV}$. The definition of the SRs and other experimental details are given in the main text.}
\end{figure}

In the two panels of Figure~\ref{fig:pheno4} we compare the $m_T^{\rm tot}$ distributions within the SM and in the benchmark~LQ model for the parameter choices $Y_{b \tau} = 2, M_{\rm LQ} = 2 \, {\rm TeV}$. The~left~(right) plot shows the results for the no~$b\hspace{0.4mm}$-tag~($b\hspace{0.4mm}$-tag) category in the $\tau_{\rm h}^+ \tau_{\rm h}^-$ final state. The~black curves represent the SM expectations of the~ DY~background taken from~\cite{CMS-PAS-HIG-21-001}, while the yellow and red histograms are the~LQ~LO and~LQ~NLO predictions obtained using our {\tt POWHEG-BOX} implementation. All results assume $138 \, {\rm fb}^{-1}$ of $pp$ data collected at $\sqrt{s} = 13 \, {\rm TeV}$. It is evident from the lower left panel that in the no~$b\hspace{0.4mm}$-tag category the NLO~LQ contribution amounts to a relative correction of only about 10\% compared to the SM~DY~background for $m_T^{\rm tot} > 1300 \, {\rm GeV}$. In the case of the $b\hspace{0.4mm}$-tag category, one instead observes from the lower right panel that in the highest $m_T^{\rm tot}$ bin with $m_T^{\rm tot} > 900 \, {\rm GeV}$ the NLO~LQ signal constitutes almost 110\% of the SM~DY~background. This again demonstrates that for third-generation scalar~LQs the sensitivity of DY~searches notably improve by demanding additional $b \hspace{0.4mm}$-jet activity. Notice that the NLO~QCD effects enhance the LO~LQ predictions in the no~$b\hspace{0.4mm}$-tag~($b\hspace{0.4mm}$-tag) category by approximately 40\% (30\%) in the highest $m_T^{\rm tot}$ bin, making higher-order QCD effects phenomenologically relevant if one wants to obtain precise predictions. On the other hand, EW and interference effects are both insignificant in the tail of the $m_T^{\rm tot}$ distribution and are therefore not shown in the~figure. 

\section{Exclusion limits}
\label{sec:limits} 

\begin{figure}[t!]
\centering
\begin{subfigure}{0.475\textwidth}
\centering
\includegraphics[width=\textwidth]{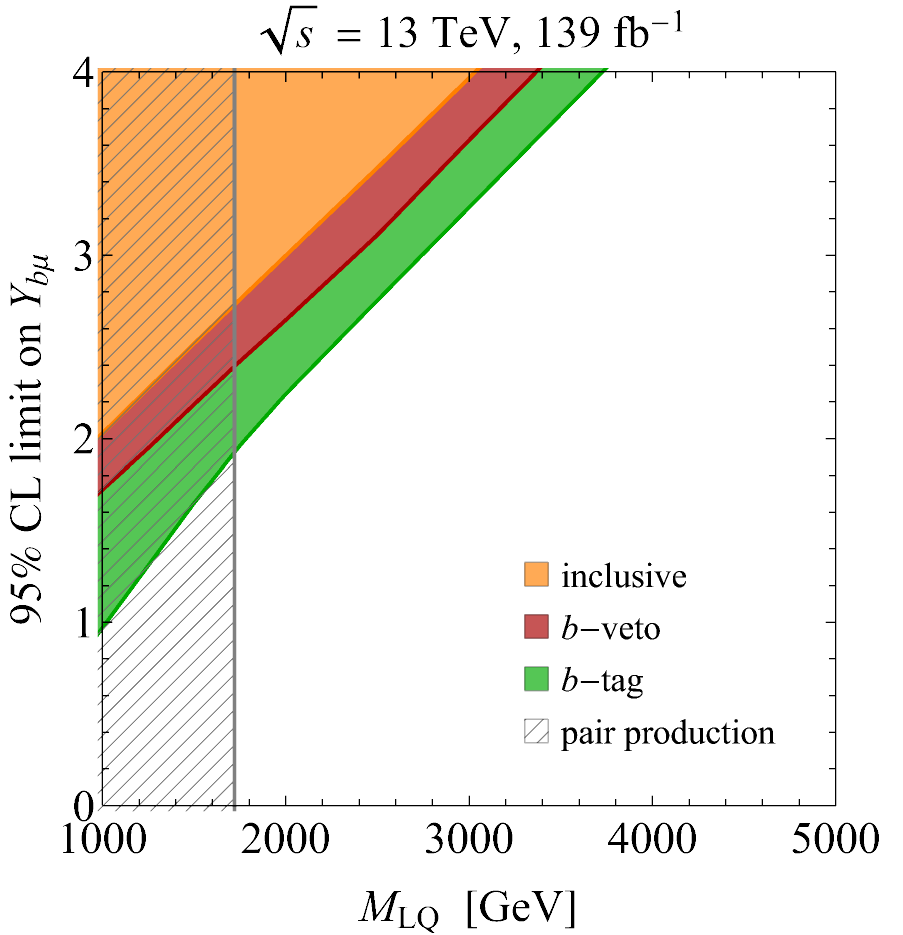}
\end{subfigure}
\quad 
\begin{subfigure}{0.475\textwidth}
\centering
\includegraphics[width=\textwidth]{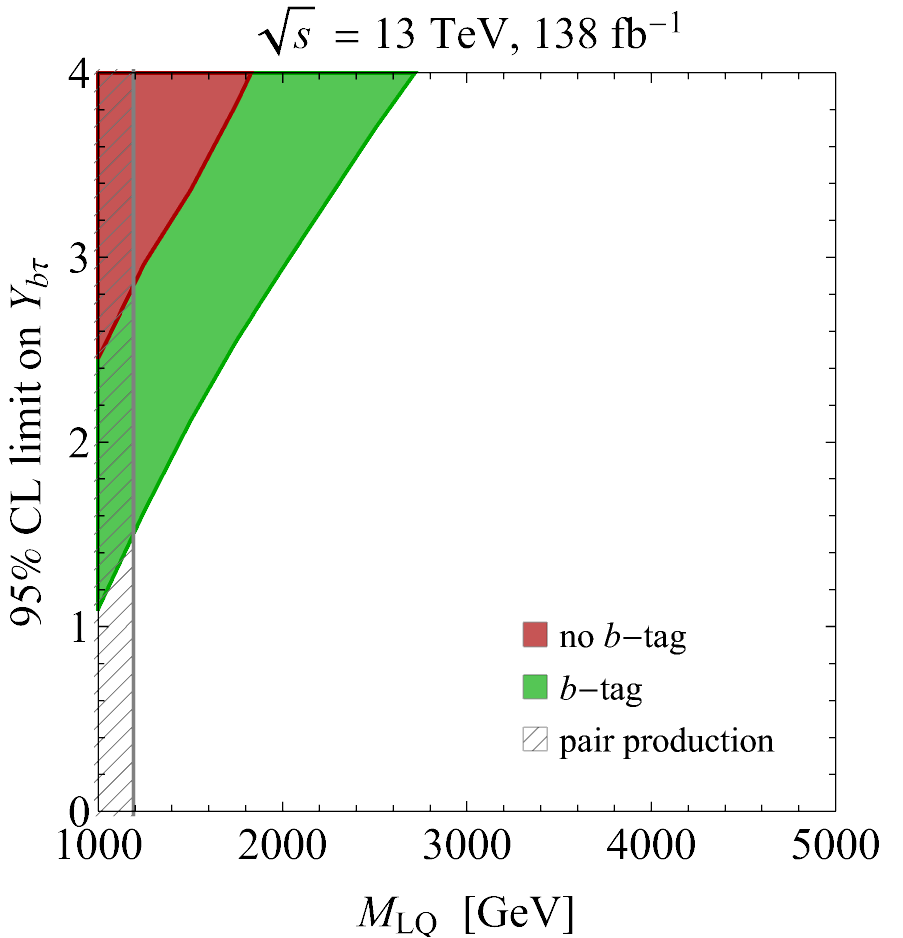}
\end{subfigure}
\vspace{2mm}
\caption{\label{fig:excl1} Left:~Comparison of the 95\%~CL constraints on the $M_{\rm LQ}\hspace{0.4mm}$--$\hspace{0.4mm} Y_{b\mu}$ plane following from different search strategies at LHC~Run~II. The~yellow,~red and~green limit corresponds to the inclusive~DY~search~\cite{ATLAS:2020yat} and the DY~analysis~\cite{ATLAS:2021mla} in the $b\hspace{0.4mm}$-veto and $b\hspace{0.4mm}$-tag category, respectively. The~hatched grey parameter space is instead excluded by the search~\cite{ATLAS:2020dsk} for strong~LQ pair production. Right:~Comparison of the 95\%~CL constraints on the $M_{\rm LQ}\hspace{0.4mm}$--$\hspace{0.4mm} Y_{b\tau}$ plane that arise from the LHC~Run~II double hadronic ditau analysis~\cite{CMS-PAS-HIG-21-001}. The~green~(red) exclusion corresponds to the no~$b\hspace{0.4mm}$-tag ($b\hspace{0.4mm}$-tag) category of the latter search, while the hatched grey parameter space is excluded by strong pair production of third-generation~LQs~\cite{ATLAS:2021jyv}. Consult the main text for additional~explanations.}
\end{figure}

On the basis of the search strategies detailed in~Section~\ref{sec:analysis}, we will now derive 95\%~confidence level~(CL) constraints on the $M_{\rm LQ}\hspace{0.4mm}$--$\hspace{0.4mm} Y_{b\mu}$ and $M_{\rm LQ}\hspace{0.4mm}$--$\hspace{0.4mm} Y_{b\tau}$ planes using the latest LHC dilepton analyses performed at LHC~Run~II. Additional exclusions limits on the parameter space of second-generation scalar~LQs can be found in Appendix~\ref{app:second}. In the left panel of Figure~\ref{fig:excl1} we show the 95\%~CL limits on the $M_{\rm LQ}\hspace{0.4mm}$--$\hspace{0.4mm} Y_{b\mu}$ parameter space. The~yellow,~red and~green bound arises from the inclusive~DY~search~\cite{ATLAS:2020yat} and the DY~analysis~\cite{ATLAS:2021mla} in the $b\hspace{0.4mm}$-veto and $b\hspace{0.4mm}$-tag category, respectively. All exclusions are based on NLO+PS predictions obtained with the {\tt POWHEG-BOX}. The hatched grey region of parameter space with $M_{\rm LQ} < 1720 \, {\rm GeV}$ is finally excluded by the search~\cite{ATLAS:2020dsk} for QCD pair production of scalar~LQs. The displayed exclusions are derived directly  from the observed model-independent upper 95\%~CL limits on the visible cross section times branching ratio provided in~\cite{ATLAS:2020yat,ATLAS:2021mla}. From the shown results it is evident that the search strategy that requires besides two OSSF muons a~$b\hspace{0.4mm}$-tag leads to the best exclusion. As explained in Section~\ref{sec:bjetlightpheno} this is to be expected because the requirement of an additional $b\hspace{0.4mm}$-tagged jet leads to a strong reduction of the signal-to-background ratio. Notice~also that for $M_{\rm LQ} \lesssim 1.7 \, {\rm TeV}$ the exclusions contour starts to~deviate from its linear behaviour. This is a consequence of the contribution associated to single-LQ production with subsequent decay of the~LQ,~cf.~the~left diagram in~Figure~\ref{fig:diagrams2}, scaling as $|Y_{b\mu}|^2$ compared to the $|Y_{b\mu}|^4$ dependence of the~squared amplitude of the $t$-channel Born-level~LQ~contribution. Another interesting feature of the results shown on the left-hand side in~Figure~\ref{fig:excl1} is that the $b\hspace{0.4mm}$-veto search performs better than the inclusive search strategy. This feature is related to the fact that the SR with $m_{\mu \mu} \in [2070, 6000] \, {\rm GeV}$ employed in~\cite{ATLAS:2020yat} is not optimised for the LQ signals studied~here. Using~the model-independent limits as a function of the minimum dimuon invariant mass~$m_{\mu \mu}^{\rm min}$,~i.e.~$m_{\mu \mu} > m_{\mu \mu}^{\rm min}$, presented in~\cite{ATLAS:2021mla} that covers lower values of~$m_{\mu \mu}^{\rm min}$ instead allows for such an optimisation and therefore leads to a stronger~bound. 

In the right panel of Figure~\ref{fig:excl1} we finally display the 95\%~CL exclusion bounds in the $M_{\rm LQ}\hspace{0.4mm}$--$\hspace{0.4mm} Y_{b\tau}$~plane that follow from the two $b\hspace{0.4mm}$-jet categories considered in the~$\tau_{\rm h}^+ \tau_{\rm h}^-$~search~\cite{CMS-PAS-HIG-21-001}. The~green and~red exclusion corresponds to the no~$b\hspace{0.4mm}$-tag and the~$b\hspace{0.4mm}$-tag category of this analysis, respectively, while the parameter space excluded by strong pair production~of~third-generation~LQs~\cite{ATLAS:2021jyv} is indicated by the hatched grey vertical band. This~search excludes $M_{\rm LQ} < 1190 \, {\rm GeV}$ at 95\%~CL. The significance  of the individual   $b\hspace{0.4mm}$-jet categories of the search~\cite{CMS-PAS-HIG-21-001} is calculated as a Poisson ratio of likelihoods modified to incorporate systematic uncertainties on the background as Gaussian constraints~\cite{Cowan:2010js}.  Our statistical analysis includes the six (three) highest $m_T^{\rm tot}$ bins in the case of the  no~$b\hspace{0.4mm}$-tag ($b\hspace{0.4mm}$-tag) category. As~for the exclusion limits on the coupling~$Y_{b\mu}$, one observes that the bound on $Y_{b\tau}$ that follows from the search with a $b\hspace{0.4mm}$-tag is more stringent than the one that derives from a strategy that vetos $b\hspace{0.4mm}$-jets. Notice lastly that as an effect of single-LQ production the slope of the exclusion arising from the $b\hspace{0.4mm}$-tag category changes at around $M_{\rm LQ} = 1.2 \, {\rm TeV}$, although this effect is less visible in the case of the coupling $Y_{b\tau}$ than for $Y_{b\mu}$. 

\section{Conclusions and outlook}
\label{sec:conclusions}

In this article we have refined the theoretical description of DY~dilepton production in scalar~LQ~models. To achieve this goal we have calculated the NLO~QCD corrections to~$pp \to \ell^+ \ell^-$ production. The actual~computation involves the evaluation of the real and virtual corrections to the~$t$-channel Born-level contribution and the calculation of resonant single-LQ production followed by the decay of the~LQ. Besides QCD corrections we have also considered the impact of virtual EW corrections and studied the size of interference effects between the~LQ~signal and the DY~SM background. These fixed-order predictions are consistently matched to a PS employing the {\tt POWHEG} method, which makes it possible to obtain a realistic exclusive description of DY~dilepton processes in scalar~LQ models at the level of hadronic events. Our {\tt POWHEG} implementation allows in particular to generate events with one additional parton from the matrix element calculation without the introduction of a merging or matching scale. Since we believe that the presented MC generator should prove useful for everyone interested in comparing accurate theory predictions to LHC data, we will make the relevant codes to simulate NLO+PS events for the $pp \to \ell^+ \ell^-$ process in scalar~LQ models of the form~(\ref{eq:SLQ}) publicly available on the {\tt POWHEG-BOX} web page~\cite{POWHEGBOX}. 

While our MC implementation can generate dilepton DY~predictions for all couplings entering the simplified~LQ Lagrangian~(\ref{eq:SLQ}), we have confined ourselves in the main part of this work to the case of $b \to \mu$ and $b \to \tau$ flavour transitions in our phenomenological analyses. The focus on these two cases is firstly motivated by the observation that in scalar~LQ models that offer an explanation of the anomalies in~semileptonic $B$ decays, the Yukawa entries~$Y_{b\mu}$ and~$Y_{b\tau}$ are necessarily the largest couplings. Second, since for $Y_{b\mu} \neq 0$ ($Y_{b\tau} \neq 0$) DY~dimuon~(ditau) production is induced at the tree level via bottom-quark fusion, initial-state radiation will always lead to a certain amount of $b\hspace{0.4mm}$-jet activity. In such cases, devising search strategies with different $b\hspace{0.4mm}$-jet categories is expected to help improve the LHC sensitivity. To illustrate the latter point, we have performed recasts of the existing LHC~Run~II~searches~\cite{ATLAS:2020yat,ATLAS:2021mla,CMS-PAS-HIG-21-001} that employ around $140 \, {\rm fb}^{-1}$ of $pp$ data collected at $\sqrt{s} = 13 \, {\rm TeV}$. In particular, we have derived the limits on the couplings $Y_{b\mu}$ and $Y_{b\tau}$ and masses of third-generation scalar~LQs from the relevant LHC searches, considering signatures with no or one $b\hspace{0.4mm}$-jet. We found that the exclusive strategies that require the presence of an additional $b\hspace{0.4mm}$-tagged jet always perform better than inclusive searches or those that veto $b\hspace{0.25mm}$-jets. The~improvement in sensitivity is particularly important in the case of the $pp \to \mu^+ \mu^-$ searches because the top and multijet background contributions to the $b\hspace{0.4mm}$-tagged sample are compared to $pp \to \tau^+ \tau^-$ less relevant. Although we have presented in our work only results for $pp \to \mu^+ \mu^-$, the latter statement applies to $pp \to e^+ e^-$ production as well. For completeness we provide the constraints on the parameter space of second-generation scalar LQs that arise from DY dilepton production in the supplementary material that can be found in Appendix~\ref{app:second}.

Let us finally add that measurements of the DY~forward-backward asymmetry~($A_{\rm FB}$) at high dilepton invariant masses such as~\cite{CMS:2022uul} might also be used to set limits on the presence of~LQs and their interactions~\cite{Raj:2016aky}. The~forward~(backward) DY~cross section thereby includes all events with $\cos \theta > 0$~($\cos \theta < 0$) where $\theta$ denotes the angle between the incoming quark and the outgoing negatively charged lepton in the Collins-Soper frame~\cite{Collins:1977iv}. At~a~$pp$~collider like the~LHC this however means that non-zero $A_{\rm FB}$ values can  only arise from the valence quarks but not the sea quarks. Since we have discussed in this work only~LQ~processes initiated by heavy-quark fusion, we have therefore not studied the constraints that arise from $A_{\rm FB}$. We however emphasise that our MC implementation is able to calculate the first-generation scalar~LQ contributions to $A_{\rm FB}$ including NLO~QCD,~EW and interference~effects. 

\acknowledgments{We thank Thomas Hahn and  Giulia Zanderighi  as well as Silvia Zanoli  for their technical support regarding {\tt LoopTools} and {\tt POWHEG-BOX}, respectively.  The Feynman diagrams shown in this article have been drawn with {\tt JaxoDraw}~\cite{Binosi:2008ig}. LS and SS are supported by the International Max Planck Research School (IMPRS) on “Elementary Particle Physics”. Partial support by the Collaborative Research Center SFB1258 is also acknowledged. UH~and~LS would like to express gratitude to the Mainz Institute for Theoretical Physics (MITP) of the Cluster of Excellence PRISMA+ (Project ID 39083149), for its hospitality and support.} 

\begin{appendix}

\section{Supplementary material}
\label{app:second}

Employing the search strategies detailed already in~Section~\ref{sec:analysis}, we present in this appendix the 95\%~CL exclusion limits on the $M_{\rm LQ}\hspace{0.4mm}$--$\hspace{0.4mm} Y_{s\mu}$ and $M_{\rm LQ}\hspace{0.4mm}$--$\hspace{0.4mm} Y_{c\tau}$ planes using the latest LHC dilepton analyses performed at LHC~Run~II. Such limits are of interest because besides the Yukawa entries $Y_{b\mu}$ and $Y_{b\tau}$  discussed in~Section~\ref{sec:limits} also $Y_{s\mu}$ and $Y_{c\tau}$  enter the predictions for $b \to s \mu^+ \mu^-$ and $b \to c \tau \nu$ in scalar~LQ models. All results displayed below are based on NLO+PS predictions obtained with our dedicated {\tt POWHEG-BOX} implementation of the interaction Lagrangian~(\ref{eq:SLQ}). Our statistical analyses employ the methodologies that have been briefly described in Section~\ref{sec:limits}.

\begin{figure}[t!]
\centering
\begin{subfigure}{0.475\textwidth}
\centering
\includegraphics[width=\textwidth]{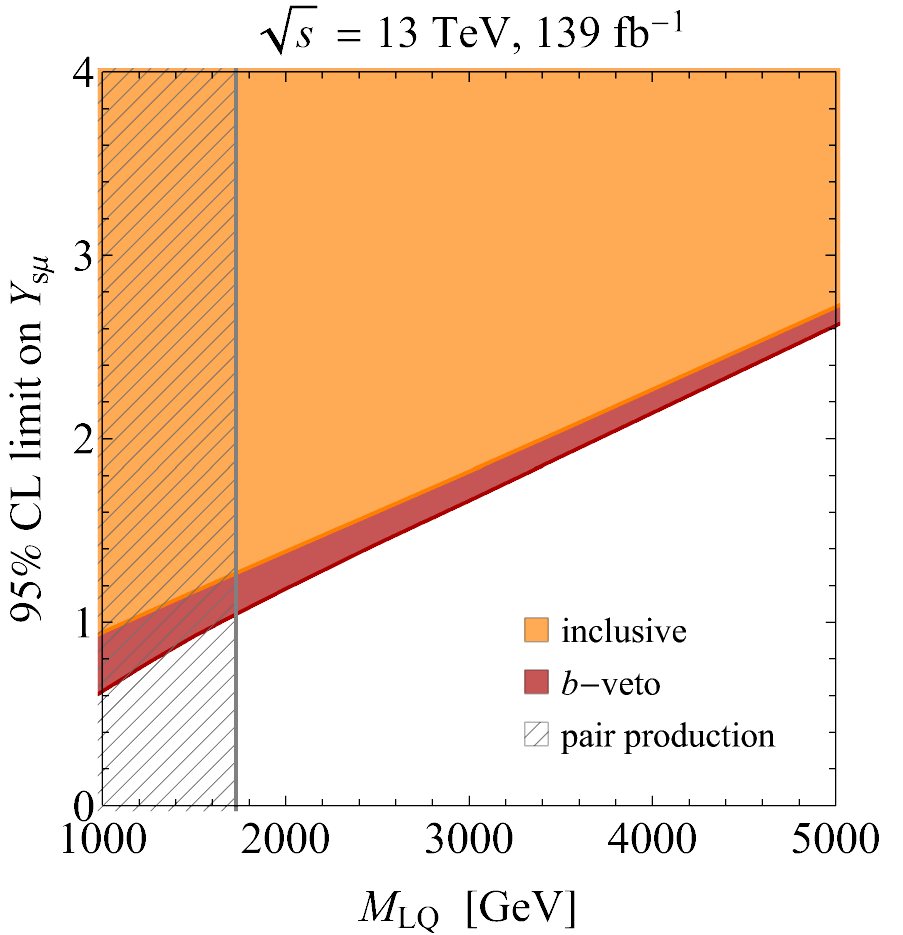}
\end{subfigure}
\quad 
\begin{subfigure}{0.475\textwidth}
\centering
\includegraphics[width=\textwidth]{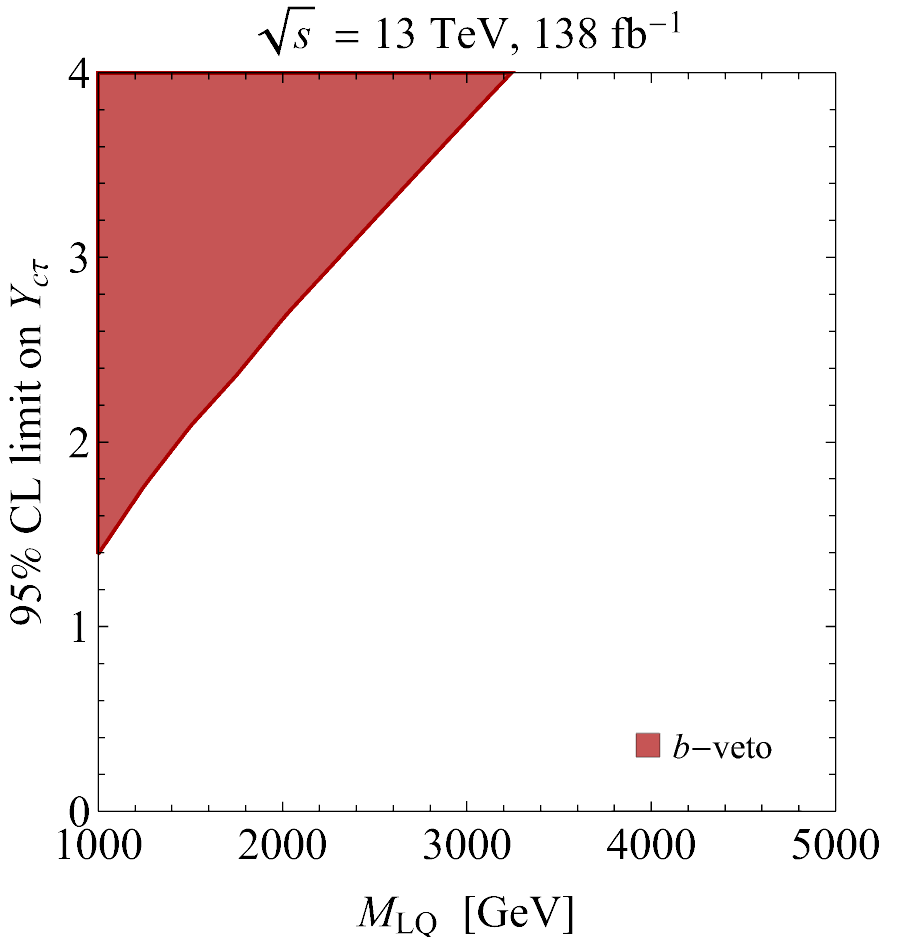}
\end{subfigure}
\vspace{2mm}
\caption{\label{fig:excl2} As~Figure~\ref{fig:excl1} but for the couplings $Y_{s\mu}$~(left panel) and $Y_{c\tau}$~(right panel). See main text for further explanations.}
\end{figure}

The~yellow and~red bound shown on the left-hand side of~Figure~\ref{fig:excl2} corresponds to the inclusive~search~\cite{ATLAS:2020yat} and the~analysis~\cite{ATLAS:2021mla} imposing a $b\hspace{0.4mm}$-veto. For comparison, we also display the parameter space with $M_{\rm LQ} < 1730 \, {\rm GeV}$ that is excluded by the search~\cite{ATLAS:2020dsk} for strong~LQ pair production as a hatched grey vertical band.  Like in the case of the coupling~$Y_{b\mu}$,~cf.~the~left panel in Figure~\ref{fig:excl1}, one sees that the exclusion following from the $b\hspace{0.4mm}$-veto search surpasses the limit that derives from the inclusive analysis. The reason is again that by choosing $m_{\mu \mu}^{\rm min}$ appropriately the sensitivity of the~$b\hspace{0.4mm}$-veto search can be improved over that of the inclusive analysis which uses a fixed and rather high value of~$m_{\mu \mu}^{\rm min}$. On~the~right in~Figure~\ref{fig:excl2} we finally present the  95\%~CL exclusion limit on the $M_{\rm LQ}\hspace{0.4mm}$--$\hspace{0.4mm} Y_{c\tau}$ plane that originates from a recast of the search with a $b\hspace{0.4mm}$-veto that has been performed in the publication~\cite{ATLAS:2021mla}. Notice that neither ATLAS nor CMS has searched for pairs of scalar~LQs decaying into light-flavour quarks and tau leptons. This explains why no bound from QCD~LQ~pair production included in the right panel of~Figure~\ref{fig:excl2}. Finally, we add that a DY ditau search that requires a $c\hspace{0.4mm}$-tag is likely to allow to strengthen the exclusion bounds on the~$M_{\rm LQ}\hspace{0.4mm}$--$\hspace{0.4mm} Y_{c\tau}$ plane compared to the limits presented in this appendix. Given~the latest advances in tagging charm quarks at the LHC~\cite{ATLAS:2019bwq,Qu:2019gqs} and the successful applications of these techniques in the recent searches for the SM Higgs boson decaying to charm-quark pairs~\cite{ATLAS:2022ers,CMS:2022psv}, we believe that OSSF dilepton searches with the requirement of an additional~$c\hspace{0.4mm}$-jet would be an interesting addition to the exotics search canon of both the ATLAS and CMS collaborations. 

\end{appendix}


\begin{thebibliography}{90}%
\makeatletter
\providecommand \@ifxundefined [1]{%
 \@ifx{#1\undefined}
}%
\providecommand \@ifnum [1]{%
 \ifnum #1\expandafter \@firstoftwo
 \else \expandafter \@secondoftwo
 \fi
}%
\providecommand \@ifx [1]{%
 \ifx #1\expandafter \@firstoftwo
 \else \expandafter \@secondoftwo
 \fi
}%
\providecommand \natexlab [1]{#1}%
\providecommand \enquote  [1]{``#1''}%
\providecommand \bibnamefont  [1]{#1}%
\providecommand \bibfnamefont [1]{#1}%
\providecommand \citenamefont [1]{#1}%
\providecommand \href@noop [0]{\@secondoftwo}%
\providecommand \href [0]{\begingroup \@sanitize@url \@href}%
\providecommand \@href[1]{\@@startlink{#1}\@@href}%
\providecommand \@@href[1]{\endgroup#1\@@endlink}%
\providecommand \@sanitize@url [0]{\catcode `\\12\catcode `\$12\catcode
  `\&12\catcode `\#12\catcode `\^12\catcode `\_12\catcode `\%12\relax}%
\providecommand \@@startlink[1]{}%
\providecommand \@@endlink[0]{}%
\providecommand \url  [0]{\begingroup\@sanitize@url \@url }%
\providecommand \@url [1]{\endgroup\@href {#1}{\urlprefix }}%
\providecommand \urlprefix  [0]{URL }%
\providecommand \Eprint [0]{\href }%
\providecommand \doibase [0]{http://dx.doi.org/}%
\providecommand \selectlanguage [0]{\@gobble}%
\providecommand \bibinfo  [0]{\@secondoftwo}%
\providecommand \bibfield  [0]{\@secondoftwo}%
\providecommand \translation [1]{[#1]}%
\providecommand \BibitemOpen [0]{}%
\providecommand \bibitemStop [0]{}%
\providecommand \bibitemNoStop [0]{.\EOS\space}%
\providecommand \EOS [0]{\spacefactor3000\relax}%
\providecommand\BibitemShut  [1]{\csname bibitem#1\endcsname}%
\let\auto@bib@innerbib\@empty
\bibitem [{\citenamefont {Aad}\ \emph {et~al.}(2020{\natexlab{a}})\citenamefont
  {Aad} \emph {et~al.}}]{ATLAS:2020zms}%
  \BibitemOpen
  \bibfield  {author} {\bibinfo {author} {\bibfnamefont {G.}~\bibnamefont
  {Aad}} \emph {et~al.} (\bibinfo {collaboration} {ATLAS}),\ }\href {\doibase
  10.1103/PhysRevLett.125.051801} {\bibfield  {journal} {\bibinfo  {journal}
  {Phys. Rev. Lett.}\ }\textbf {\bibinfo {volume} {125}},\ \bibinfo {pages}
  {051801} (\bibinfo {year} {2020}{\natexlab{a}})},\ \Eprint
  {http://arxiv.org/abs/2002.12223} {arXiv:2002.12223 [hep-ex]}\BibitemShut
  {NoStop}%
\bibitem [{\citenamefont {Aad}\ \emph {et~al.}(2020{\natexlab{b}})\citenamefont
  {Aad} \emph {et~al.}}]{ATLAS:2020yat}%
  \BibitemOpen
  \bibfield  {author} {\bibinfo {author} {\bibfnamefont {G.}~\bibnamefont
  {Aad}} \emph {et~al.} (\bibinfo {collaboration} {ATLAS}),\ }\href {\doibase
  10.1007/JHEP11(2020)005} {\bibfield  {journal} {\bibinfo  {journal} {JHEP}\
  }\textbf {\bibinfo {volume} {11}},\ \bibinfo {pages} {005} (\bibinfo {year}
  {2020}{\natexlab{b}})},\ \bibinfo {note} {[Erratum: JHEP {\bf 04}, 142 (2021)]},\
  \Eprint {http://arxiv.org/abs/2006.12946} {arXiv:2006.12946 [hep-ex]}\BibitemShut {NoStop}%
\bibitem [{\citenamefont {Sirunyan}\ \emph {et~al.}(2021)\citenamefont
  {Sirunyan} \emph {et~al.}}]{CMS:2021ctt}%
  \BibitemOpen
  \bibfield  {author} {\bibinfo {author} {\bibfnamefont {A.~M.}\ \bibnamefont
  {Sirunyan}} \emph {et~al.} (\bibinfo {collaboration} {CMS}),\ }\href
  {\doibase 10.1007/JHEP07(2021)208} {\bibfield  {journal} {\bibinfo  {journal}
  {JHEP}\ }\textbf {\bibinfo {volume} {07}},\ \bibinfo {pages} {208} (\bibinfo
  {year} {2021})},\ \Eprint {http://arxiv.org/abs/2103.02708} {arXiv:2103.02708
  [hep-ex]}\BibitemShut {NoStop}%
\bibitem [{\citenamefont {Aad}\ \emph {et~al.}(2021{\natexlab{a}})\citenamefont
  {Aad} \emph {et~al.}}]{ATLAS:2021mla}%
  \BibitemOpen
  \bibfield  {author} {\bibinfo {author} {\bibfnamefont {G.}~\bibnamefont
  {Aad}} \emph {et~al.} (\bibinfo {collaboration} {ATLAS}),\ }\href {\doibase
  10.1103/PhysRevLett.127.141801} {\bibfield  {journal} {\bibinfo  {journal}
  {Phys. Rev. Lett.}\ }\textbf {\bibinfo {volume} {127}},\ \bibinfo {pages}
  {141801} (\bibinfo {year} {2021}{\natexlab{a}})},\ \Eprint
  {http://arxiv.org/abs/2105.13847} {arXiv:2105.13847 [hep-ex]}\BibitemShut
  {NoStop}%
\bibitem [{\citenamefont {Tumasyan}\ \emph
  {et~al.}(2022{\natexlab{a}})\citenamefont {Tumasyan} \emph
  {et~al.}}]{CMS:2022uul}%
  \BibitemOpen
  \bibfield  {author} {\bibinfo {author} {\bibfnamefont {A.}~\bibnamefont
  {Tumasyan}} \emph {et~al.} (\bibinfo {collaboration} {CMS}),\ }\href@noop {}
  {\  (\bibinfo {year} {2022}{\natexlab{a}})},\ \Eprint
  {http://arxiv.org/abs/2202.12327} {arXiv:2202.12327 [hep-ex]}\BibitemShut
  {NoStop}%
\bibitem [{CMS(2022)}]{CMS-PAS-HIG-21-001}%
  \BibitemOpen
  \href {https://cds.cern.ch/record/2803739} {\emph {\bibinfo {title}
  {{Searches for additional Higgs bosons and vector leptoquarks in $\tau\tau$
  final states in proton-proton collisions at $\sqrt{s}=13~\mathrm{TeV}$}}}},\
 \bibinfo  {institution} {CERN},\ \bibinfo
  {address} {Geneva},\ \bibinfo {year} {2022}\BibitemShut {NoStop}%
\bibitem [{\citenamefont {Faroughy}\ \emph {et~al.}(2017)\citenamefont
  {Faroughy}, \citenamefont {Greljo},\ and\ \citenamefont
  {Kamenik}}]{Faroughy:2016osc}%
  \BibitemOpen
  \bibfield  {author} {\bibinfo {author} {\bibfnamefont {D.~A.}\ \bibnamefont
  {Faroughy}}, \bibinfo {author} {\bibfnamefont {A.}~\bibnamefont {Greljo}}, \
  and\ \bibinfo {author} {\bibfnamefont {J.~F.}\ \bibnamefont {Kamenik}},\
  }\href {\doibase 10.1016/j.physletb.2016.11.011} {\bibfield  {journal}
  {\bibinfo  {journal} {Phys. Lett. B}\ }\textbf {\bibinfo {volume} {764}},\
  \bibinfo {pages} {126} (\bibinfo {year} {2017})},\ \Eprint
  {http://arxiv.org/abs/1609.07138} {arXiv:1609.07138 [hep-ph]}\BibitemShut
  {NoStop}%
\bibitem [{\citenamefont {Raj}(2017)}]{Raj:2016aky}%
  \BibitemOpen
  \bibfield  {author} {\bibinfo {author} {\bibfnamefont {N.}~\bibnamefont
  {Raj}},\ }\href {\doibase 10.1103/PhysRevD.95.015011} {\bibfield  {journal}
  {\bibinfo  {journal} {Phys. Rev. D}\ }\textbf {\bibinfo {volume} {95}},\
  \bibinfo {pages} {015011} (\bibinfo {year} {2017})},\ \Eprint
  {http://arxiv.org/abs/1610.03795} {arXiv:1610.03795 [hep-ph]}\BibitemShut
  {NoStop}%
\bibitem [{\citenamefont {Greljo}\ and\ \citenamefont
  {Marzocca}(2017)}]{Greljo:2017vvb}%
  \BibitemOpen
  \bibfield  {author} {\bibinfo {author} {\bibfnamefont {A.}~\bibnamefont
  {Greljo}}\ and\ \bibinfo {author} {\bibfnamefont {D.}~\bibnamefont
  {Marzocca}},\ }\href {\doibase 10.1140/epjc/s10052-017-5119-8} {\bibfield
  {journal} {\bibinfo  {journal} {Eur. Phys. J. C}\ }\textbf {\bibinfo {volume}
  {77}},\ \bibinfo {pages} {548} (\bibinfo {year} {2017})},\ \Eprint
  {http://arxiv.org/abs/1704.09015} {arXiv:1704.09015 [hep-ph]}\BibitemShut
  {NoStop}%
\bibitem [{\citenamefont {Allanach}\ \emph {et~al.}(2018)\citenamefont
  {Allanach}, \citenamefont {Gripaios},\ and\ \citenamefont
  {You}}]{Allanach:2017bta}%
  \BibitemOpen
  \bibfield  {author} {\bibinfo {author} {\bibfnamefont {B.~C.}\ \bibnamefont
  {Allanach}}, \bibinfo {author} {\bibfnamefont {B.}~\bibnamefont {Gripaios}},
  \ and\ \bibinfo {author} {\bibfnamefont {T.}~\bibnamefont {You}},\ }\href
  {\doibase 10.1007/JHEP03(2018)021} {\bibfield  {journal} {\bibinfo  {journal}
  {JHEP}\ }\textbf {\bibinfo {volume} {03}},\ \bibinfo {pages} {021} (\bibinfo
  {year} {2018})},\ \Eprint {http://arxiv.org/abs/1710.06363} {arXiv:1710.06363
  [hep-ph]}\BibitemShut {NoStop}%
\bibitem [{\citenamefont {Dor\v{s}ner}\ and\ \citenamefont
  {Greljo}(2018)}]{Dorsner:2018ynv}%
  \BibitemOpen
  \bibfield  {author} {\bibinfo {author} {\bibfnamefont {I.}~\bibnamefont
  {Dor\v{s}ner}}\ and\ \bibinfo {author} {\bibfnamefont {A.}~\bibnamefont
  {Greljo}},\ }\href {\doibase 10.1007/JHEP05(2018)126} {\bibfield  {journal}
  {\bibinfo  {journal} {JHEP}\ }\textbf {\bibinfo {volume} {05}},\ \bibinfo
  {pages} {126} (\bibinfo {year} {2018})},\ \Eprint
  {http://arxiv.org/abs/1801.07641} {arXiv:1801.07641 [hep-ph]}\BibitemShut
  {NoStop}%
\bibitem [{\citenamefont {Afik}\ \emph {et~al.}(2018)\citenamefont {Afik},
  \citenamefont {Cohen}, \citenamefont {Gozani}, \citenamefont {Kajomovitz},\
  and\ \citenamefont {Rozen}}]{Afik:2018nlr}%
  \BibitemOpen
  \bibfield  {author} {\bibinfo {author} {\bibfnamefont {Y.}~\bibnamefont
  {Afik}}, \bibinfo {author} {\bibfnamefont {J.}~\bibnamefont {Cohen}},
  \bibinfo {author} {\bibfnamefont {E.}~\bibnamefont {Gozani}}, \bibinfo
  {author} {\bibfnamefont {E.}~\bibnamefont {Kajomovitz}}, \ and\ \bibinfo
  {author} {\bibfnamefont {Y.}~\bibnamefont {Rozen}},\ }\href {\doibase
  10.1007/JHEP08(2018)056} {\bibfield  {journal} {\bibinfo  {journal} {JHEP}\
  }\textbf {\bibinfo {volume} {08}},\ \bibinfo {pages} {056} (\bibinfo {year}
  {2018})},\ \Eprint {http://arxiv.org/abs/1805.11402} {arXiv:1805.11402
  [hep-ph]}\BibitemShut {NoStop}%
\bibitem [{\citenamefont {Bansal}\ \emph {et~al.}(2018)\citenamefont {Bansal},
  \citenamefont {Capdevilla}, \citenamefont {Delgado}, \citenamefont {Kolda},
  \citenamefont {Martin},\ and\ \citenamefont {Raj}}]{Bansal:2018eha}%
  \BibitemOpen
  \bibfield  {author} {\bibinfo {author} {\bibfnamefont {S.}~\bibnamefont
  {Bansal}}, \bibinfo {author} {\bibfnamefont {R.~M.}\ \bibnamefont
  {Capdevilla}}, \bibinfo {author} {\bibfnamefont {A.}~\bibnamefont {Delgado}},
  \bibinfo {author} {\bibfnamefont {C.}~\bibnamefont {Kolda}}, \bibinfo
  {author} {\bibfnamefont {A.}~\bibnamefont {Martin}}, \ and\ \bibinfo {author}
  {\bibfnamefont {N.}~\bibnamefont {Raj}},\ }\href {\doibase
  10.1103/PhysRevD.98.015037} {\bibfield  {journal} {\bibinfo  {journal} {Phys.
  Rev. D}\ }\textbf {\bibinfo {volume} {98}},\ \bibinfo {pages} {015037}
  (\bibinfo {year} {2018})},\ \Eprint {http://arxiv.org/abs/1806.02370}
  {arXiv:1806.02370 [hep-ph]}\BibitemShut {NoStop}%
\bibitem [{\citenamefont {Allanach}\ \emph {et~al.}(2019)\citenamefont
  {Allanach}, \citenamefont {Corbett}, \citenamefont {Dolan},\ and\
  \citenamefont {You}}]{Allanach:2018odd}%
  \BibitemOpen
  \bibfield  {author} {\bibinfo {author} {\bibfnamefont {B.~C.}\ \bibnamefont
  {Allanach}}, \bibinfo {author} {\bibfnamefont {T.}~\bibnamefont {Corbett}},
  \bibinfo {author} {\bibfnamefont {M.~J.}\ \bibnamefont {Dolan}}, \ and\
  \bibinfo {author} {\bibfnamefont {T.}~\bibnamefont {You}},\ }\href {\doibase
  10.1007/JHEP03(2019)137} {\bibfield  {journal} {\bibinfo  {journal} {JHEP}\
  }\textbf {\bibinfo {volume} {03}},\ \bibinfo {pages} {137} (\bibinfo {year}
  {2019})},\ \Eprint {http://arxiv.org/abs/1810.02166} {arXiv:1810.02166
  [hep-ph]}\BibitemShut {NoStop}%
\bibitem [{\citenamefont {Schmaltz}\ and\ \citenamefont
  {Zhong}(2019)}]{Schmaltz:2018nls}%
  \BibitemOpen
  \bibfield  {author} {\bibinfo {author} {\bibfnamefont {M.}~\bibnamefont
  {Schmaltz}}\ and\ \bibinfo {author} {\bibfnamefont {Y.-M.}\ \bibnamefont
  {Zhong}},\ }\href {\doibase 10.1007/JHEP01(2019)132} {\bibfield  {journal}
  {\bibinfo  {journal} {JHEP}\ }\textbf {\bibinfo {volume} {01}},\ \bibinfo
  {pages} {132} (\bibinfo {year} {2019})},\ \Eprint
  {http://arxiv.org/abs/1810.10017} {arXiv:1810.10017 [hep-ph]}\BibitemShut
  {NoStop}%
\bibitem [{\citenamefont {Mandal}\ \emph {et~al.}(2019)\citenamefont {Mandal},
  \citenamefont {Mitra},\ and\ \citenamefont {Raz}}]{Mandal:2018kau}%
  \BibitemOpen
  \bibfield  {author} {\bibinfo {author} {\bibfnamefont {T.}~\bibnamefont
  {Mandal}}, \bibinfo {author} {\bibfnamefont {S.}~\bibnamefont {Mitra}}, \
  and\ \bibinfo {author} {\bibfnamefont {S.}~\bibnamefont {Raz}},\ }\href
  {\doibase 10.1103/PhysRevD.99.055028} {\bibfield  {journal} {\bibinfo
  {journal} {Phys. Rev. D}\ }\textbf {\bibinfo {volume} {99}},\ \bibinfo
  {pages} {055028} (\bibinfo {year} {2019})},\ \Eprint
  {http://arxiv.org/abs/1811.03561} {arXiv:1811.03561 [hep-ph]}\BibitemShut
  {NoStop}%
\bibitem [{\citenamefont {Baker}\ \emph {et~al.}(2019)\citenamefont {Baker},
  \citenamefont {Fuentes-Mart\'\i{}n}, \citenamefont {Isidori},\ and\
  \citenamefont {K\"onig}}]{Baker:2019sli}%
  \BibitemOpen
  \bibfield  {author} {\bibinfo {author} {\bibfnamefont {M.~J.}\ \bibnamefont
  {Baker}}, \bibinfo {author} {\bibfnamefont {J.}~\bibnamefont
  {Fuentes-Mart\'\i{}n}}, \bibinfo {author} {\bibfnamefont {G.}~\bibnamefont
  {Isidori}}, \ and\ \bibinfo {author} {\bibfnamefont {M.}~\bibnamefont
  {K\"onig}},\ }\href {\doibase 10.1140/epjc/s10052-019-6853-x} {\bibfield
  {journal} {\bibinfo  {journal} {Eur. Phys. J. C}\ }\textbf {\bibinfo {volume}
  {79}},\ \bibinfo {pages} {334} (\bibinfo {year} {2019})},\ \Eprint
  {http://arxiv.org/abs/1901.10480} {arXiv:1901.10480 [hep-ph]}\BibitemShut
  {NoStop}%
\bibitem [{\citenamefont {Choudhury}\ \emph {et~al.}(2019)\citenamefont
  {Choudhury}, \citenamefont {Kumar},\ and\ \citenamefont
  {Kundu}}]{Choudhury:2019ucz}%
  \BibitemOpen
  \bibfield  {author} {\bibinfo {author} {\bibfnamefont {D.}~\bibnamefont
  {Choudhury}}, \bibinfo {author} {\bibfnamefont {N.}~\bibnamefont {Kumar}}, \
  and\ \bibinfo {author} {\bibfnamefont {A.}~\bibnamefont {Kundu}},\ }\href
  {\doibase 10.1103/PhysRevD.100.075001} {\bibfield  {journal} {\bibinfo
  {journal} {Phys. Rev. D}\ }\textbf {\bibinfo {volume} {100}},\ \bibinfo
  {pages} {075001} (\bibinfo {year} {2019})},\ \Eprint
  {http://arxiv.org/abs/1905.07982} {arXiv:1905.07982 [hep-ph]}\BibitemShut
  {NoStop}%
\bibitem [{\citenamefont {Angelescu}\ \emph {et~al.}(2020)\citenamefont
  {Angelescu}, \citenamefont {Faroughy},\ and\ \citenamefont
  {Sumensari}}]{Angelescu:2020uug}%
  \BibitemOpen
  \bibfield  {author} {\bibinfo {author} {\bibfnamefont {A.}~\bibnamefont
  {Angelescu}}, \bibinfo {author} {\bibfnamefont {D.~A.}\ \bibnamefont
  {Faroughy}}, \ and\ \bibinfo {author} {\bibfnamefont {O.}~\bibnamefont
  {Sumensari}},\ }\href {\doibase 10.1140/epjc/s10052-020-8210-5} {\bibfield
  {journal} {\bibinfo  {journal} {Eur. Phys. J. C}\ }\textbf {\bibinfo {volume}
  {80}},\ \bibinfo {pages} {641} (\bibinfo {year} {2020})},\ \Eprint
  {http://arxiv.org/abs/2002.05684} {arXiv:2002.05684 [hep-ph]}\BibitemShut
  {NoStop}%
\bibitem [{\citenamefont {Bhaskar}\ \emph {et~al.}(2021)\citenamefont
  {Bhaskar}, \citenamefont {Das}, \citenamefont {Mandal}, \citenamefont
  {Mitra},\ and\ \citenamefont {Neeraj}}]{Bhaskar:2021pml}%
  \BibitemOpen
  \bibfield  {author} {\bibinfo {author} {\bibfnamefont {A.}~\bibnamefont
  {Bhaskar}}, \bibinfo {author} {\bibfnamefont {D.}~\bibnamefont {Das}},
  \bibinfo {author} {\bibfnamefont {T.}~\bibnamefont {Mandal}}, \bibinfo
  {author} {\bibfnamefont {S.}~\bibnamefont {Mitra}}, \ and\ \bibinfo {author}
  {\bibfnamefont {C.}~\bibnamefont {Neeraj}},\ }\href {\doibase
  10.1103/PhysRevD.104.035016} {\bibfield  {journal} {\bibinfo  {journal}
  {Phys. Rev. D}\ }\textbf {\bibinfo {volume} {104}},\ \bibinfo {pages}
  {035016} (\bibinfo {year} {2021})},\ \Eprint
  {http://arxiv.org/abs/2101.12069} {arXiv:2101.12069 [hep-ph]}\BibitemShut
  {NoStop}%
\bibitem [{\citenamefont {Crivellin}\ \emph
  {et~al.}(2021{\natexlab{a}})\citenamefont {Crivellin}, \citenamefont
  {Manzari},\ and\ \citenamefont {Montull}}]{Crivellin:2021rbf}%
  \BibitemOpen
  \bibfield  {author} {\bibinfo {author} {\bibfnamefont {A.}~\bibnamefont
  {Crivellin}}, \bibinfo {author} {\bibfnamefont {C.~A.}\ \bibnamefont
  {Manzari}}, \ and\ \bibinfo {author} {\bibfnamefont {M.}~\bibnamefont
  {Montull}},\ }\href {\doibase 10.1103/PhysRevD.104.115016} {\bibfield
  {journal} {\bibinfo  {journal} {Phys. Rev. D}\ }\textbf {\bibinfo {volume}
  {104}},\ \bibinfo {pages} {115016} (\bibinfo {year} {2021}{\natexlab{a}})},\
  \Eprint {http://arxiv.org/abs/2103.12003} {arXiv:2103.12003 [hep-ph]}
 \BibitemShut {NoStop}%
\bibitem [{\citenamefont {Cornella}\ \emph {et~al.}(2021)\citenamefont
  {Cornella}, \citenamefont {Faroughy}, \citenamefont {Fuentes-Martin},
  \citenamefont {Isidori},\ and\ \citenamefont {Neubert}}]{Cornella:2021sby}%
  \BibitemOpen
  \bibfield  {author} {\bibinfo {author} {\bibfnamefont {C.}~\bibnamefont
  {Cornella}}, \bibinfo {author} {\bibfnamefont {D.~A.}\ \bibnamefont
  {Faroughy}}, \bibinfo {author} {\bibfnamefont {J.}~\bibnamefont
  {Fuentes-Martin}}, \bibinfo {author} {\bibfnamefont {G.}~\bibnamefont
  {Isidori}}, \ and\ \bibinfo {author} {\bibfnamefont {M.}~\bibnamefont
  {Neubert}},\ }\href {\doibase 10.1007/JHEP08(2021)050} {\bibfield  {journal}
  {\bibinfo  {journal} {JHEP}\ }\textbf {\bibinfo {volume} {08}},\ \bibinfo
  {pages} {050} (\bibinfo {year} {2021})},\ \Eprint
  {http://arxiv.org/abs/2103.16558} {arXiv:2103.16558 [hep-ph]}\BibitemShut
  {NoStop}%
\bibitem [{\citenamefont {Crivellin}\ \emph
  {et~al.}(2021{\natexlab{b}})\citenamefont {Crivellin}, \citenamefont
  {M\"uller},\ and\ \citenamefont {Schnell}}]{Crivellin:2021egp}%
  \BibitemOpen
  \bibfield  {author} {\bibinfo {author} {\bibfnamefont {A.}~\bibnamefont
  {Crivellin}}, \bibinfo {author} {\bibfnamefont {D.}~\bibnamefont {M\"uller}},
  \ and\ \bibinfo {author} {\bibfnamefont {L.}~\bibnamefont {Schnell}},\ }\href
  {\doibase 10.1103/PhysRevD.103.115023} {\bibfield  {journal} {\bibinfo
  {journal} {Phys. Rev. D}\ }\textbf {\bibinfo {volume} {103}},\ \bibinfo
  {pages} {115023} (\bibinfo {year} {2021}{\natexlab{b}})},\ \Eprint
  {http://arxiv.org/abs/2104.06417} {arXiv:2104.06417 [hep-ph]}\BibitemShut
  {NoStop}%
\bibitem [{\citenamefont {Crivellin}\ \emph
  {et~al.}(2021{\natexlab{c}})\citenamefont {Crivellin}, \citenamefont
  {Hoferichter}, \citenamefont {Kirk}, \citenamefont {Manzari},\ and\
  \citenamefont {Schnell}}]{Crivellin:2021bkd}%
  \BibitemOpen
  \bibfield  {author} {\bibinfo {author} {\bibfnamefont {A.}~\bibnamefont
  {Crivellin}}, \bibinfo {author} {\bibfnamefont {M.}~\bibnamefont
  {Hoferichter}}, \bibinfo {author} {\bibfnamefont {M.}~\bibnamefont {Kirk}},
  \bibinfo {author} {\bibfnamefont {C.~A.}\ \bibnamefont {Manzari}}, \ and\
  \bibinfo {author} {\bibfnamefont {L.}~\bibnamefont {Schnell}},\ }\href
  {\doibase 10.1007/JHEP10(2021)221} {\bibfield  {journal} {\bibinfo  {journal}
  {JHEP}\ }\textbf {\bibinfo {volume} {10}},\ \bibinfo {pages} {221} (\bibinfo
  {year} {2021}{\natexlab{c}})},\ \Eprint {http://arxiv.org/abs/2107.13569}
  {arXiv:2107.13569 [hep-ph]}\BibitemShut {NoStop}%
\bibitem [{\citenamefont {Garland}\ \emph {et~al.}(2022)\citenamefont
  {Garland}, \citenamefont {J\"ager}, \citenamefont {Khosa},\ and\
  \citenamefont {Kvedarait\.{e}}}]{Garland:2021ghw}%
  \BibitemOpen
  \bibfield  {author} {\bibinfo {author} {\bibfnamefont {B.}~\bibnamefont
  {Garland}}, \bibinfo {author} {\bibfnamefont {S.}~\bibnamefont {J\"ager}},
  \bibinfo {author} {\bibfnamefont {C.~K.}\ \bibnamefont {Khosa}}, \ and\
  \bibinfo {author} {\bibfnamefont {S.}~\bibnamefont {Kvedarait\.{e}}},\ }\href
  {\doibase 10.1103/PhysRevD.105.115017} {\bibfield  {journal} {\bibinfo
  {journal} {Phys. Rev. D}\ }\textbf {\bibinfo {volume} {105}},\ \bibinfo
  {pages} {115017} (\bibinfo {year} {2022})},\ \Eprint
  {http://arxiv.org/abs/2112.05127} {arXiv:2112.05127 [hep-ph]}\BibitemShut
  {NoStop}%
\bibitem [{\citenamefont {Crivellin}\ \emph {et~al.}(2022)\citenamefont
  {Crivellin}, \citenamefont {Fuks},\ and\ \citenamefont
  {Schnell}}]{Crivellin:2022mff}%
  \BibitemOpen
  \bibfield  {author} {\bibinfo {author} {\bibfnamefont {A.}~\bibnamefont
  {Crivellin}}, \bibinfo {author} {\bibfnamefont {B.}~\bibnamefont {Fuks}}, \
  and\ \bibinfo {author} {\bibfnamefont {L.}~\bibnamefont {Schnell}},\
  }\href@noop {} {\  (\bibinfo {year} {2022})},\ \Eprint
  {http://arxiv.org/abs/2203.10111} {arXiv:2203.10111 [hep-ph]}\BibitemShut
  {NoStop}%
\bibitem [{\citenamefont {Azatov}\ \emph {et~al.}(2022)\citenamefont {Azatov},
  \citenamefont {Garosi}, \citenamefont {Greljo}, \citenamefont {Marzocca},
  \citenamefont {Salko},\ and\ \citenamefont {Trifinopoulos}}]{Azatov:2022itm}%
  \BibitemOpen
  \bibfield  {author} {\bibinfo {author} {\bibfnamefont {A.}~\bibnamefont
  {Azatov}}, \bibinfo {author} {\bibfnamefont {F.}~\bibnamefont {Garosi}},
  \bibinfo {author} {\bibfnamefont {A.}~\bibnamefont {Greljo}}, \bibinfo
  {author} {\bibfnamefont {D.}~\bibnamefont {Marzocca}}, \bibinfo {author}
  {\bibfnamefont {J.}~\bibnamefont {Salko}}, \ and\ \bibinfo {author}
  {\bibfnamefont {S.}~\bibnamefont {Trifinopoulos}},\ }\href@noop {} {\
  (\bibinfo {year} {2022})},\ \Eprint {http://arxiv.org/abs/2205.13552}
  {arXiv:2205.13552 [hep-ph]}\BibitemShut {NoStop}%
\bibitem [{\citenamefont {Lees}\ \emph {et~al.}(2012)\citenamefont {Lees} \emph
  {et~al.}}]{BaBar:2012obs}%
  \BibitemOpen
  \bibfield  {author} {\bibinfo {author} {\bibfnamefont {J.~P.}\ \bibnamefont
  {Lees}} \emph {et~al.} (\bibinfo {collaboration} {BaBar}),\ }\href {\doibase
  10.1103/PhysRevLett.109.101802} {\bibfield  {journal} {\bibinfo  {journal}
  {Phys. Rev. Lett.}\ }\textbf {\bibinfo {volume} {109}},\ \bibinfo {pages}
  {101802} (\bibinfo {year} {2012})},\ \Eprint {http://arxiv.org/abs/1205.5442}
  {arXiv:1205.5442 [hep-ex]}\BibitemShut {NoStop}%
\bibitem [{\citenamefont {Lees}\ \emph {et~al.}(2013)\citenamefont {Lees} \emph
  {et~al.}}]{BaBar:2013mob}%
  \BibitemOpen
  \bibfield  {author} {\bibinfo {author} {\bibfnamefont {J.~P.}\ \bibnamefont
  {Lees}} \emph {et~al.} (\bibinfo {collaboration} {BaBar}),\ }\href {\doibase
  10.1103/PhysRevD.88.072012} {\bibfield  {journal} {\bibinfo  {journal} {Phys.
  Rev. D}\ }\textbf {\bibinfo {volume} {88}},\ \bibinfo {pages} {072012}
  (\bibinfo {year} {2013})},\ \Eprint {http://arxiv.org/abs/1303.0571}
  {arXiv:1303.0571 [hep-ex]}\BibitemShut {NoStop}%
\bibitem [{\citenamefont {Aaij}\ \emph {et~al.}(2015)\citenamefont {Aaij} \emph
  {et~al.}}]{LHCb:2015gmp}%
  \BibitemOpen
  \bibfield  {author} {\bibinfo {author} {\bibfnamefont {R.}~\bibnamefont
  {Aaij}} \emph {et~al.} (\bibinfo {collaboration} {LHCb}),\ }\href {\doibase
  10.1103/PhysRevLett.115.111803} {\bibfield  {journal} {\bibinfo  {journal}
  {Phys. Rev. Lett.}\ }\textbf {\bibinfo {volume} {115}},\ \bibinfo {pages}
  {111803} (\bibinfo {year} {2015})},\ \bibinfo {note} {[Erratum: Phys. Rev.
  Lett. {\bf 115}, 159901 (2015)]},\ \Eprint {http://arxiv.org/abs/1506.08614}
  {arXiv:1506.08614 [hep-ex]}\BibitemShut {NoStop}%
\bibitem [{\citenamefont {Aaij}\ \emph
  {et~al.}(2018{\natexlab{a}})\citenamefont {Aaij} \emph
  {et~al.}}]{LHCb:2017smo}%
  \BibitemOpen
  \bibfield  {author} {\bibinfo {author} {\bibfnamefont {R.}~\bibnamefont
  {Aaij}} \emph {et~al.} (\bibinfo {collaboration} {LHCb}),\ }\href {\doibase
  10.1103/PhysRevLett.120.171802} {\bibfield  {journal} {\bibinfo  {journal}
  {Phys. Rev. Lett.}\ }\textbf {\bibinfo {volume} {120}},\ \bibinfo {pages}
  {171802} (\bibinfo {year} {2018}{\natexlab{a}})},\ \Eprint
  {http://arxiv.org/abs/1708.08856} {arXiv:1708.08856 [hep-ex]}\BibitemShut
  {NoStop}%
\bibitem [{\citenamefont {Aaij}\ \emph
  {et~al.}(2018{\natexlab{b}})\citenamefont {Aaij} \emph
  {et~al.}}]{LHCb:2017rln}%
  \BibitemOpen
  \bibfield  {author} {\bibinfo {author} {\bibfnamefont {R.}~\bibnamefont
  {Aaij}} \emph {et~al.} (\bibinfo {collaboration} {LHCb}),\ }\href {\doibase
  10.1103/PhysRevD.97.072013} {\bibfield  {journal} {\bibinfo  {journal} {Phys.
  Rev. D}\ }\textbf {\bibinfo {volume} {97}},\ \bibinfo {pages} {072013}
  (\bibinfo {year} {2018}{\natexlab{b}})},\ \Eprint
  {http://arxiv.org/abs/1711.02505} {arXiv:1711.02505 [hep-ex]}\BibitemShut
  {NoStop}%
\bibitem [{\citenamefont {Abdesselam}\ \emph {et~al.}(2019)\citenamefont
  {Abdesselam} \emph {et~al.}}]{Belle:2019gij}%
  \BibitemOpen
  \bibfield  {author} {\bibinfo {author} {\bibfnamefont {A.}~\bibnamefont
  {Abdesselam}} \emph {et~al.} (\bibinfo {collaboration} {Belle}),\ }\href@noop
  {} {\  (\bibinfo {year} {2019})},\ \Eprint {http://arxiv.org/abs/1904.08794}
  {arXiv:1904.08794 [hep-ex]}\BibitemShut {NoStop}%
\bibitem [{\citenamefont {Aaij}\ \emph {et~al.}(2017)\citenamefont {Aaij} \emph
  {et~al.}}]{LHCb:2017avl}%
  \BibitemOpen
  \bibfield  {author} {\bibinfo {author} {\bibfnamefont {R.}~\bibnamefont
  {Aaij}} \emph {et~al.} (\bibinfo {collaboration} {LHCb}),\ }\href {\doibase
  10.1007/JHEP08(2017)055} {\bibfield  {journal} {\bibinfo  {journal} {JHEP}\
  }\textbf {\bibinfo {volume} {08}},\ \bibinfo {pages} {055} (\bibinfo {year}
  {2017})},\ \Eprint {http://arxiv.org/abs/1705.05802} {arXiv:1705.05802
  [hep-ex]}\BibitemShut {NoStop}%
\bibitem [{\citenamefont {Aaij}\ \emph {et~al.}(2019)\citenamefont {Aaij} \emph
  {et~al.}}]{LHCb:2019hip}%
  \BibitemOpen
  \bibfield  {author} {\bibinfo {author} {\bibfnamefont {R.}~\bibnamefont
  {Aaij}} \emph {et~al.} (\bibinfo {collaboration} {LHCb}),\ }\href {\doibase
  10.1103/PhysRevLett.122.191801} {\bibfield  {journal} {\bibinfo  {journal}
  {Phys. Rev. Lett.}\ }\textbf {\bibinfo {volume} {122}},\ \bibinfo {pages}
  {191801} (\bibinfo {year} {2019})},\ \Eprint
  {http://arxiv.org/abs/1903.09252} {arXiv:1903.09252 [hep-ex]}\BibitemShut
  {NoStop}%
\bibitem [{\citenamefont {Abdesselam}\ \emph {et~al.}(2021)\citenamefont
  {Abdesselam} \emph {et~al.}}]{Belle:2019oag}%
  \BibitemOpen
  \bibfield  {author} {\bibinfo {author} {\bibfnamefont {A.}~\bibnamefont
  {Abdesselam}} \emph {et~al.} (\bibinfo {collaboration} {Belle}),\ }\href
  {\doibase 10.1103/PhysRevLett.126.161801} {\bibfield  {journal} {\bibinfo
  {journal} {Phys. Rev. Lett.}\ }\textbf {\bibinfo {volume} {126}},\ \bibinfo
  {pages} {161801} (\bibinfo {year} {2021})},\ \Eprint
  {http://arxiv.org/abs/1904.02440} {arXiv:1904.02440 [hep-ex]}\BibitemShut
  {NoStop}%
\bibitem [{\citenamefont {Choudhury}\ \emph {et~al.}(2021)\citenamefont
  {Choudhury} \emph {et~al.}}]{BELLE:2019xld}%
  \BibitemOpen
  \bibfield  {author} {\bibinfo {author} {\bibfnamefont {S.}~\bibnamefont
  {Choudhury}} \emph {et~al.} (\bibinfo {collaboration} {Belle}),\ }\href
  {\doibase 10.1007/JHEP03(2021)105} {\bibfield  {journal} {\bibinfo  {journal}
  {JHEP}\ }\textbf {\bibinfo {volume} {03}},\ \bibinfo {pages} {105} (\bibinfo
  {year} {2021})},\ \Eprint {http://arxiv.org/abs/1908.01848} {arXiv:1908.01848
  [hep-ex]}\BibitemShut {NoStop}%
\bibitem [{\citenamefont {Aaij}\ \emph {et~al.}(2022)\citenamefont {Aaij} \emph
  {et~al.}}]{LHCb:2021trn}%
  \BibitemOpen
  \bibfield  {author} {\bibinfo {author} {\bibfnamefont {R.}~\bibnamefont
  {Aaij}} \emph {et~al.} (\bibinfo {collaboration} {LHCb}),\ }\href {\doibase
  10.1038/s41567-021-01478-8} {\bibfield  {journal} {\bibinfo  {journal}
  {Nature Phys.}\ }\textbf {\bibinfo {volume} {18}},\ \bibinfo {pages} {277}
  (\bibinfo {year} {2022})},\ \Eprint {http://arxiv.org/abs/2103.11769}
  {arXiv:2103.11769 [hep-ex]}\BibitemShut {NoStop}%
\bibitem [{\citenamefont {Altmannshofer}\ \emph {et~al.}(2017)\citenamefont
  {Altmannshofer}, \citenamefont {Bhupal~Dev},\ and\ \citenamefont
  {Soni}}]{Altmannshofer:2017poe}%
  \BibitemOpen
  \bibfield  {author} {\bibinfo {author} {\bibfnamefont {W.}~\bibnamefont
  {Altmannshofer}}, \bibinfo {author} {\bibfnamefont {P.~S.}\ \bibnamefont
  {Bhupal~Dev}}, \ and\ \bibinfo {author} {\bibfnamefont {A.}~\bibnamefont
  {Soni}},\ }\href {\doibase 10.1103/PhysRevD.96.095010} {\bibfield  {journal}
  {\bibinfo  {journal} {Phys. Rev. D}\ }\textbf {\bibinfo {volume} {96}},\
  \bibinfo {pages} {095010} (\bibinfo {year} {2017})},\ \Eprint
  {http://arxiv.org/abs/1704.06659} {arXiv:1704.06659 [hep-ph]}\BibitemShut
  {NoStop}%
\bibitem [{\citenamefont {Iguro}\ and\ \citenamefont
  {Tobe}(2017)}]{Iguro:2017ysu}%
  \BibitemOpen
  \bibfield  {author} {\bibinfo {author} {\bibfnamefont {S.}~\bibnamefont
  {Iguro}}\ and\ \bibinfo {author} {\bibfnamefont {K.}~\bibnamefont {Tobe}},\
  }\href {\doibase 10.1016/j.nuclphysb.2017.10.014} {\bibfield  {journal}
  {\bibinfo  {journal} {Nucl. Phys. B}\ }\textbf {\bibinfo {volume} {925}},\
  \bibinfo {pages} {560} (\bibinfo {year} {2017})},\ \Eprint
  {http://arxiv.org/abs/1708.06176} {arXiv:1708.06176 [hep-ph]}\BibitemShut
  {NoStop}%
\bibitem [{\citenamefont {Abdullah}\ \emph {et~al.}(2018)\citenamefont
  {Abdullah}, \citenamefont {Calle}, \citenamefont {Dutta}, \citenamefont
  {Fl\'orez},\ and\ \citenamefont {Restrepo}}]{Abdullah:2018ets}%
  \BibitemOpen
  \bibfield  {author} {\bibinfo {author} {\bibfnamefont {M.}~\bibnamefont
  {Abdullah}}, \bibinfo {author} {\bibfnamefont {J.}~\bibnamefont {Calle}},
  \bibinfo {author} {\bibfnamefont {B.}~\bibnamefont {Dutta}}, \bibinfo
  {author} {\bibfnamefont {A.}~\bibnamefont {Fl\'orez}}, \ and\ \bibinfo
  {author} {\bibfnamefont {D.}~\bibnamefont {Restrepo}},\ }\href {\doibase
  10.1103/PhysRevD.98.055016} {\bibfield  {journal} {\bibinfo  {journal} {Phys.
  Rev. D}\ }\textbf {\bibinfo {volume} {98}},\ \bibinfo {pages} {055016}
  (\bibinfo {year} {2018})},\ \Eprint {http://arxiv.org/abs/1805.01869}
  {arXiv:1805.01869 [hep-ph]}\BibitemShut {NoStop}%
\bibitem [{\citenamefont {Marzocca}\ \emph {et~al.}(2020)\citenamefont
  {Marzocca}, \citenamefont {Min},\ and\ \citenamefont
  {Son}}]{Marzocca:2020ueu}%
  \BibitemOpen
  \bibfield  {author} {\bibinfo {author} {\bibfnamefont {D.}~\bibnamefont
  {Marzocca}}, \bibinfo {author} {\bibfnamefont {U.}~\bibnamefont {Min}}, \
  and\ \bibinfo {author} {\bibfnamefont {M.}~\bibnamefont {Son}},\ }\href
  {\doibase 10.1007/JHEP12(2020)035} {\bibfield  {journal} {\bibinfo  {journal}
  {JHEP}\ }\textbf {\bibinfo {volume} {12}},\ \bibinfo {pages} {035} (\bibinfo
  {year} {2020})},\ \Eprint {http://arxiv.org/abs/2008.07541} {arXiv:2008.07541
  [hep-ph]}\BibitemShut {NoStop}%
\bibitem [{\citenamefont {Endo}\ \emph {et~al.}(2022)\citenamefont {Endo},
  \citenamefont {Iguro}, \citenamefont {Kitahara}, \citenamefont {Takeuchi},\
  and\ \citenamefont {Watanabe}}]{Endo:2021lhi}%
  \BibitemOpen
  \bibfield  {author} {\bibinfo {author} {\bibfnamefont {M.}~\bibnamefont
  {Endo}}, \bibinfo {author} {\bibfnamefont {S.}~\bibnamefont {Iguro}},
  \bibinfo {author} {\bibfnamefont {T.}~\bibnamefont {Kitahara}}, \bibinfo
  {author} {\bibfnamefont {M.}~\bibnamefont {Takeuchi}}, \ and\ \bibinfo
  {author} {\bibfnamefont {R.}~\bibnamefont {Watanabe}},\ }\href {\doibase
  10.1007/JHEP02(2022)106} {\bibfield  {journal} {\bibinfo  {journal} {JHEP}\
  }\textbf {\bibinfo {volume} {02}},\ \bibinfo {pages} {106} (\bibinfo {year}
  {2022})},\ \Eprint {http://arxiv.org/abs/2111.04748} {arXiv:2111.04748
  [hep-ph]}\BibitemShut {NoStop}%
\bibitem [{\citenamefont {Kr{\"a}mer}\ \emph {et~al.}(1997)\citenamefont
  {Kr{\"a}mer}, \citenamefont {Plehn}, \citenamefont {Spira},\ and\
  \citenamefont {Zerwas}}]{Kramer:1997hh}%
  \BibitemOpen
  \bibfield  {author} {\bibinfo {author} {\bibfnamefont {M.}~\bibnamefont
  {Kr{\"a}mer}}, \bibinfo {author} {\bibfnamefont {T.}~\bibnamefont {Plehn}},
  \bibinfo {author} {\bibfnamefont {M.}~\bibnamefont {Spira}}, \ and\ \bibinfo
  {author} {\bibfnamefont {P.~M.}\ \bibnamefont {Zerwas}},\ }\href {\doibase
  10.1103/PhysRevLett.79.341} {\bibfield  {journal} {\bibinfo  {journal} {Phys.
  Rev. Lett.}\ }\textbf {\bibinfo {volume} {79}},\ \bibinfo {pages} {341}
  (\bibinfo {year} {1997})},\ \Eprint {http://arxiv.org/abs/hep-ph/9704322}
  {arXiv:hep-ph/9704322}\BibitemShut {NoStop}%
\bibitem [{\citenamefont {Kr{\"a}mer}\ \emph {et~al.}(2005)\citenamefont
  {Kr{\"a}mer}, \citenamefont {Plehn}, \citenamefont {Spira},\ and\
  \citenamefont {Zerwas}}]{Kramer:2004df}%
  \BibitemOpen
  \bibfield  {author} {\bibinfo {author} {\bibfnamefont {M.}~\bibnamefont
  {Kr{\"a}mer}}, \bibinfo {author} {\bibfnamefont {T.}~\bibnamefont {Plehn}},
  \bibinfo {author} {\bibfnamefont {M.}~\bibnamefont {Spira}}, \ and\ \bibinfo
  {author} {\bibfnamefont {P.~M.}\ \bibnamefont {Zerwas}},\ }\href {\doibase
  10.1103/PhysRevD.71.057503} {\bibfield  {journal} {\bibinfo  {journal} {Phys.
  Rev. D}\ }\textbf {\bibinfo {volume} {71}},\ \bibinfo {pages} {057503}
  (\bibinfo {year} {2005})},\ \Eprint {http://arxiv.org/abs/hep-ph/0411038}
  {arXiv:hep-ph/0411038}\BibitemShut {NoStop}%
\bibitem [{\citenamefont {Hammett}\ and\ \citenamefont
  {Ross}(2015)}]{Hammett:2015sea}%
  \BibitemOpen
  \bibfield  {author} {\bibinfo {author} {\bibfnamefont {J.~B.}\ \bibnamefont
  {Hammett}}\ and\ \bibinfo {author} {\bibfnamefont {D.~A.}\ \bibnamefont
  {Ross}},\ }\href {\doibase 10.1007/JHEP07(2015)148} {\bibfield  {journal}
  {\bibinfo  {journal} {JHEP}\ }\textbf {\bibinfo {volume} {07}},\ \bibinfo
  {pages} {148} (\bibinfo {year} {2015})},\ \Eprint
  {http://arxiv.org/abs/1501.06719} {arXiv:1501.06719 [hep-ph]}\BibitemShut
  {NoStop}%
\bibitem [{\citenamefont {Mandal}\ \emph {et~al.}(2016)\citenamefont {Mandal},
  \citenamefont {Mitra},\ and\ \citenamefont {Seth}}]{Mandal:2015lca}%
  \BibitemOpen
  \bibfield  {author} {\bibinfo {author} {\bibfnamefont {T.}~\bibnamefont
  {Mandal}}, \bibinfo {author} {\bibfnamefont {S.}~\bibnamefont {Mitra}}, \
  and\ \bibinfo {author} {\bibfnamefont {S.}~\bibnamefont {Seth}},\ }\href
  {\doibase 10.1103/PhysRevD.93.035018} {\bibfield  {journal} {\bibinfo
  {journal} {Phys. Rev. D}\ }\textbf {\bibinfo {volume} {93}},\ \bibinfo
  {pages} {035018} (\bibinfo {year} {2016})},\ \Eprint
  {http://arxiv.org/abs/1506.07369} {arXiv:1506.07369 [hep-ph]}\BibitemShut
  {NoStop}%
\bibitem [{\citenamefont {Borschensky}\ \emph {et~al.}(2020)\citenamefont
  {Borschensky}, \citenamefont {Fuks}, \citenamefont {Kulesza},\ and\
  \citenamefont {Schwartl\"ander}}]{Borschensky:2020hot}%
  \BibitemOpen
  \bibfield  {author} {\bibinfo {author} {\bibfnamefont {C.}~\bibnamefont
  {Borschensky}}, \bibinfo {author} {\bibfnamefont {B.}~\bibnamefont {Fuks}},
  \bibinfo {author} {\bibfnamefont {A.}~\bibnamefont {Kulesza}}, \ and\
  \bibinfo {author} {\bibfnamefont {D.}~\bibnamefont {Schwartl\"ander}},\
  }\href {\doibase 10.1103/PhysRevD.101.115017} {\bibfield  {journal} {\bibinfo
   {journal} {Phys. Rev. D}\ }\textbf {\bibinfo {volume} {101}},\ \bibinfo
  {pages} {115017} (\bibinfo {year} {2020})},\ \Eprint
  {http://arxiv.org/abs/2002.08971} {arXiv:2002.08971 [hep-ph]}\BibitemShut
  {NoStop}%
\bibitem [{\citenamefont {Buonocore}\ \emph
  {et~al.}(2020{\natexlab{a}})\citenamefont {Buonocore}, \citenamefont
  {Haisch}, \citenamefont {Nason}, \citenamefont {Tramontano},\ and\
  \citenamefont {Zanderighi}}]{Buonocore:2020erb}%
  \BibitemOpen
  \bibfield  {author} {\bibinfo {author} {\bibfnamefont {L.}~\bibnamefont
  {Buonocore}}, \bibinfo {author} {\bibfnamefont {U.}~\bibnamefont {Haisch}},
  \bibinfo {author} {\bibfnamefont {P.}~\bibnamefont {Nason}}, \bibinfo
  {author} {\bibfnamefont {F.}~\bibnamefont {Tramontano}}, \ and\ \bibinfo
  {author} {\bibfnamefont {G.}~\bibnamefont {Zanderighi}},\ }\href {\doibase
  10.1103/PhysRevLett.125.231804} {\bibfield  {journal} {\bibinfo  {journal}
  {Phys. Rev. Lett.}\ }\textbf {\bibinfo {volume} {125}},\ \bibinfo {pages}
  {231804} (\bibinfo {year} {2020}{\natexlab{a}})},\ \Eprint
  {http://arxiv.org/abs/2005.06475} {arXiv:2005.06475 [hep-ph]}\BibitemShut
  {NoStop}%
\bibitem [{\citenamefont {Buonocore}\ \emph
  {et~al.}(2020{\natexlab{b}})\citenamefont {Buonocore}, \citenamefont {Nason},
  \citenamefont {Tramontano},\ and\ \citenamefont
  {Zanderighi}}]{Buonocore:2020nai}%
  \BibitemOpen
  \bibfield  {author} {\bibinfo {author} {\bibfnamefont {L.}~\bibnamefont
  {Buonocore}}, \bibinfo {author} {\bibfnamefont {P.}~\bibnamefont {Nason}},
  \bibinfo {author} {\bibfnamefont {F.}~\bibnamefont {Tramontano}}, \ and\
  \bibinfo {author} {\bibfnamefont {G.}~\bibnamefont {Zanderighi}},\ }\href
  {\doibase 10.1007/JHEP08(2020)019} {\bibfield  {journal} {\bibinfo  {journal}
  {JHEP}\ }\textbf {\bibinfo {volume} {08}},\ \bibinfo {pages} {019} (\bibinfo
  {year} {2020}{\natexlab{b}})},\ \Eprint {http://arxiv.org/abs/2005.06477}
  {arXiv:2005.06477 [hep-ph]}\BibitemShut {NoStop}%
\bibitem [{\citenamefont {Greljo}\ and\ \citenamefont
  {Selimovic}(2021)}]{Greljo:2020tgv}%
  \BibitemOpen
  \bibfield  {author} {\bibinfo {author} {\bibfnamefont {A.}~\bibnamefont
  {Greljo}}\ and\ \bibinfo {author} {\bibfnamefont {N.}~\bibnamefont
  {Selimovic}},\ }\href {\doibase 10.1007/JHEP03(2021)279} {\bibfield
  {journal} {\bibinfo  {journal} {JHEP}\ }\textbf {\bibinfo {volume} {03}},\
  \bibinfo {pages} {279} (\bibinfo {year} {2021})},\ \Eprint
  {http://arxiv.org/abs/2012.02092} {arXiv:2012.02092 [hep-ph]}\BibitemShut
  {NoStop}%
\bibitem [{\citenamefont {Haisch}\ and\ \citenamefont
  {Polesello}(2021)}]{Haisch:2020xjd}%
  \BibitemOpen
  \bibfield  {author} {\bibinfo {author} {\bibfnamefont {U.}~\bibnamefont
  {Haisch}}\ and\ \bibinfo {author} {\bibfnamefont {G.}~\bibnamefont
  {Polesello}},\ }\href {\doibase 10.1007/JHEP05(2021)057} {\bibfield
  {journal} {\bibinfo  {journal} {JHEP}\ }\textbf {\bibinfo {volume} {05}},\
  \bibinfo {pages} {057} (\bibinfo {year} {2021})},\ \Eprint
  {http://arxiv.org/abs/2012.11474} {arXiv:2012.11474 [hep-ph]}\BibitemShut
  {NoStop}%
\bibitem [{\citenamefont {Borschensky}\ \emph {et~al.}(2022)\citenamefont
  {Borschensky}, \citenamefont {Fuks}, \citenamefont {Kulesza},\ and\
  \citenamefont {Schwartl\"ander}}]{Borschensky:2021hbo}%
  \BibitemOpen
  \bibfield  {author} {\bibinfo {author} {\bibfnamefont {C.}~\bibnamefont
  {Borschensky}}, \bibinfo {author} {\bibfnamefont {B.}~\bibnamefont {Fuks}},
  \bibinfo {author} {\bibfnamefont {A.}~\bibnamefont {Kulesza}}, \ and\
  \bibinfo {author} {\bibfnamefont {D.}~\bibnamefont {Schwartl\"ander}},\
  }\href {\doibase 10.1007/JHEP02(2022)157} {\bibfield  {journal} {\bibinfo
  {journal} {JHEP}\ }\textbf {\bibinfo {volume} {02}},\ \bibinfo {pages} {157}
  (\bibinfo {year} {2022})},\ \Eprint {http://arxiv.org/abs/2108.11404}
  {arXiv:2108.11404 [hep-ph]}\BibitemShut {NoStop}%
\bibitem [{\citenamefont {Alves}\ \emph {et~al.}(2019)\citenamefont {Alves},
  \citenamefont {Eboli}, \citenamefont {Grilli Di~Cortona},\ and\ \citenamefont
  {Moreira}}]{Alves:2018krf}%
  \BibitemOpen
  \bibfield  {author} {\bibinfo {author} {\bibfnamefont {A.}~\bibnamefont
  {Alves}}, \bibinfo {author} {\bibfnamefont {O.~J. P.}\ \bibnamefont
  {{\'E}boli}}, \bibinfo {author} {\bibfnamefont {G.}~\bibnamefont {Grilli
  Di~Cortona}}, \ and\ \bibinfo {author} {\bibfnamefont {R.~R.}\ \bibnamefont
  {Moreira}},\ }\href {\doibase 10.1103/PhysRevD.99.095005} {\bibfield
  {journal} {\bibinfo  {journal} {Phys. Rev. D}\ }\textbf {\bibinfo {volume}
  {99}},\ \bibinfo {pages} {095005} (\bibinfo {year} {2019})},\ \Eprint
  {http://arxiv.org/abs/1812.08632} {arXiv:1812.08632 [hep-ph]}\BibitemShut
  {NoStop}%
\bibitem [{\citenamefont {Nason}(2004)}]{Nason:2004rx}%
  \BibitemOpen
  \bibfield  {author} {\bibinfo {author} {\bibfnamefont {P.}~\bibnamefont
  {Nason}},\ }\href {\doibase 10.1088/1126-6708/2004/11/040} {\bibfield
  {journal} {\bibinfo  {journal} {JHEP}\ }\textbf {\bibinfo {volume} {11}},\
  \bibinfo {pages} {040} (\bibinfo {year} {2004})},\ \Eprint
  {http://arxiv.org/abs/hep-ph/0409146} {arXiv:hep-ph/0409146}\BibitemShut
  {NoStop}%
\bibitem [{\citenamefont {Frixione}\ \emph {et~al.}(2007)\citenamefont
  {Frixione}, \citenamefont {Nason},\ and\ \citenamefont
  {Oleari}}]{Frixione:2007vw}%
  \BibitemOpen
  \bibfield  {author} {\bibinfo {author} {\bibfnamefont {S.}~\bibnamefont
  {Frixione}}, \bibinfo {author} {\bibfnamefont {P.}~\bibnamefont {Nason}}, \
  and\ \bibinfo {author} {\bibfnamefont {C.}~\bibnamefont {Oleari}},\ }\href
  {\doibase 10.1088/1126-6708/2007/11/070} {\bibfield  {journal} {\bibinfo
  {journal} {JHEP}\ }\textbf {\bibinfo {volume} {11}},\ \bibinfo {pages} {070}
  (\bibinfo {year} {2007})},\ \Eprint {http://arxiv.org/abs/0709.2092}
  {arXiv:0709.2092 [hep-ph]}\BibitemShut {NoStop}%
\bibitem [{\citenamefont {Alioli}\ \emph {et~al.}(2010)\citenamefont {Alioli},
  \citenamefont {Nason}, \citenamefont {Oleari},\ and\ \citenamefont
  {Re}}]{Alioli:2010xd}%
  \BibitemOpen
  \bibfield  {author} {\bibinfo {author} {\bibfnamefont {S.}~\bibnamefont
  {Alioli}}, \bibinfo {author} {\bibfnamefont {P.}~\bibnamefont {Nason}},
  \bibinfo {author} {\bibfnamefont {C.}~\bibnamefont {Oleari}}, \ and\ \bibinfo
  {author} {\bibfnamefont {E.}~\bibnamefont {Re}},\ }\href {\doibase
  10.1007/JHEP06(2010)043} {\bibfield  {journal} {\bibinfo  {journal} {JHEP}\
  }\textbf {\bibinfo {volume} {06}},\ \bibinfo {pages} {043} (\bibinfo {year}
  {2010})},\ \Eprint {http://arxiv.org/abs/1002.2581} {arXiv:1002.2581
  [hep-ph]}\BibitemShut {NoStop}%
\bibitem [{\citenamefont {Pati}\ and\ \citenamefont
  {Salam}(1974)}]{Pati:1974yy}%
  \BibitemOpen
  \bibfield  {author} {\bibinfo {author} {\bibfnamefont {J.~C.}\ \bibnamefont
  {Pati}}\ and\ \bibinfo {author} {\bibfnamefont {A.}~\bibnamefont {Salam}},\
  }\href {\doibase 10.1103/PhysRevD.10.275} {\bibfield  {journal} {\bibinfo
  {journal} {Phys. Rev. D}\ }\textbf {\bibinfo {volume} {10}},\ \bibinfo
  {pages} {275} (\bibinfo {year} {1974})},\ \bibinfo {note} {[Erratum: Phys.
  Rev. D {\bf{11}}, 703 (1975)]}\BibitemShut {NoStop}%
\bibitem [{\citenamefont {Pati}\ and\ \citenamefont
  {Salam}(1973)}]{Pati:1973uk}%
  \BibitemOpen
  \bibfield  {author} {\bibinfo {author} {\bibfnamefont {J.~C.}\ \bibnamefont
  {Pati}}\ and\ \bibinfo {author} {\bibfnamefont {A.}~\bibnamefont {Salam}},\
  }\href {\doibase 10.1103/PhysRevD.8.1240} {\bibfield  {journal} {\bibinfo
  {journal} {Phys. Rev. D}\ }\textbf {\bibinfo {volume} {8}},\ \bibinfo {pages}
  {1240} (\bibinfo {year} {1973})}\BibitemShut {NoStop}%
\bibitem [{\citenamefont {Buchm{\"u}ller}\ \emph {et~al.}(1987)\citenamefont
  {Buchm{\"u}ller}, \citenamefont {R{\"u}ckl},\ and\ \citenamefont
  {Wyler}}]{Buchmuller:1986zs}%
  \BibitemOpen
  \bibfield  {author} {\bibinfo {author} {\bibfnamefont {W.}~\bibnamefont
  {Buchm{\"u}ller}}, \bibinfo {author} {\bibfnamefont {R.}~\bibnamefont
  {R{\"u}ckl}}, \ and\ \bibinfo {author} {\bibfnamefont {D.}~\bibnamefont
  {Wyler}},\ }\href {\doibase 10.1016/0370-2693(87)90637-X} {\bibfield
  {journal} {\bibinfo  {journal} {Phys. Lett. B}\ }\textbf {\bibinfo {volume}
  {191}},\ \bibinfo {pages} {442} (\bibinfo {year} {1987})},\ \bibinfo {note}
  {[Erratum: Phys. Lett. B {\bf{448}}, 320 (1999)]}\BibitemShut {NoStop}%
\bibitem [{\citenamefont {Dor\v{s}ner}\ \emph {et~al.}(2016)\citenamefont
  {Dor\v{s}ner}, \citenamefont {Fajfer}, \citenamefont {Greljo}, \citenamefont
  {Kamenik},\ and\ \citenamefont {Ko\v{s}nik}}]{Dorsner:2016wpm}%
  \BibitemOpen
  \bibfield  {author} {\bibinfo {author} {\bibfnamefont {I.}~\bibnamefont
  {Dor\v{s}ner}}, \bibinfo {author} {\bibfnamefont {S.}~\bibnamefont {Fajfer}},
  \bibinfo {author} {\bibfnamefont {A.}~\bibnamefont {Greljo}}, \bibinfo
  {author} {\bibfnamefont {J.~F.}\ \bibnamefont {Kamenik}}, \ and\ \bibinfo
  {author} {\bibfnamefont {N.}~\bibnamefont {Ko\v{s}nik}},\ }\href {\doibase
  10.1016/j.physrep.2016.06.001} {\bibfield  {journal} {\bibinfo  {journal}
  {Phys. Rept.}\ }\textbf {\bibinfo {volume} {641}},\ \bibinfo {pages} {1}
  (\bibinfo {year} {2016})},\ \Eprint {http://arxiv.org/abs/1603.04993}
  {arXiv:1603.04993 [hep-ph]}\BibitemShut {NoStop}%
\bibitem [{\citenamefont {Alloul}\ \emph {et~al.}(2014)\citenamefont {Alloul},
  \citenamefont {Christensen}, \citenamefont {Degrande}, \citenamefont {Duhr},\
  and\ \citenamefont {Fuks}}]{Alloul:2013bka}%
  \BibitemOpen
  \bibfield  {author} {\bibinfo {author} {\bibfnamefont {A.}~\bibnamefont
  {Alloul}}, \bibinfo {author} {\bibfnamefont {N.~D.}\ \bibnamefont
  {Christensen}}, \bibinfo {author} {\bibfnamefont {C.}~\bibnamefont
  {Degrande}}, \bibinfo {author} {\bibfnamefont {C.}~\bibnamefont {Duhr}}, \
  and\ \bibinfo {author} {\bibfnamefont {B.}~\bibnamefont {Fuks}},\ }\href
  {\doibase 10.1016/j.cpc.2014.04.012} {\bibfield  {journal} {\bibinfo
  {journal} {Comput. Phys. Commun.}\ }\textbf {\bibinfo {volume} {185}},\
  \bibinfo {pages} {2250} (\bibinfo {year} {2014})},\ \Eprint
  {http://arxiv.org/abs/1310.1921} {arXiv:1310.1921 [hep-ph]}\BibitemShut
  {NoStop}%
\bibitem [{\citenamefont {Hahn}(2001)}]{Hahn:2000kx}%
  \BibitemOpen
  \bibfield  {author} {\bibinfo {author} {\bibfnamefont {T.}~\bibnamefont
  {Hahn}},\ }\href {\doibase 10.1016/S0010-4655(01)00290-9} {\bibfield
  {journal} {\bibinfo  {journal} {Comput. Phys. Commun.}\ }\textbf {\bibinfo
  {volume} {140}},\ \bibinfo {pages} {418} (\bibinfo {year} {2001})},\ \Eprint
  {http://arxiv.org/abs/hep-ph/0012260} {arXiv:hep-ph/0012260}\BibitemShut
  {NoStop}%
\bibitem [{\citenamefont {Hahn}\ \emph {et~al.}(2016)\citenamefont {Hahn},
  \citenamefont {Pa\ss{}ehr},\ and\ \citenamefont
  {Schappacher}}]{Hahn:2016ebn}%
  \BibitemOpen
  \bibfield  {author} {\bibinfo {author} {\bibfnamefont {T.}~\bibnamefont
  {Hahn}}, \bibinfo {author} {\bibfnamefont {S.}~\bibnamefont {Pa\ss{}ehr}}, \
  and\ \bibinfo {author} {\bibfnamefont {C.}~\bibnamefont {Schappacher}},\
  }\href {\doibase 10.1088/1742-6596/762/1/012065} {\bibfield  {journal}
  {\bibinfo  {journal} {PoS}\ }\textbf {\bibinfo {volume} {LL2016}},\ \bibinfo
  {pages} {068} (\bibinfo {year} {2016})},\ \Eprint
  {http://arxiv.org/abs/1604.04611} {arXiv:1604.04611 [hep-ph]}\BibitemShut
  {NoStop}%
\bibitem [{\citenamefont {Hahn}\ and\ \citenamefont
  {Perez-Victoria}(1999)}]{Hahn:1998yk}%
  \BibitemOpen
  \bibfield  {author} {\bibinfo {author} {\bibfnamefont {T.}~\bibnamefont
  {Hahn}}\ and\ \bibinfo {author} {\bibfnamefont {M.}~\bibnamefont
  {Perez-Victoria}},\ }\href {\doibase 10.1016/S0010-4655(98)00173-8}
  {\bibfield  {journal} {\bibinfo  {journal} {Comput. Phys. Commun.}\ }\textbf
  {\bibinfo {volume} {118}},\ \bibinfo {pages} {153} (\bibinfo {year}
  {1999})},\ \Eprint {http://arxiv.org/abs/hep-ph/9807565}
  {arXiv:hep-ph/9807565}\BibitemShut {NoStop}%
\bibitem [{\citenamefont {Patel}(2015)}]{Patel:2015tea}%
  \BibitemOpen
  \bibfield  {author} {\bibinfo {author} {\bibfnamefont {H.~H.}\ \bibnamefont
  {Patel}},\ }\href {\doibase 10.1016/j.cpc.2015.08.017} {\bibfield  {journal}
  {\bibinfo  {journal} {Comput. Phys. Commun.}\ }\textbf {\bibinfo {volume}
  {197}},\ \bibinfo {pages} {276} (\bibinfo {year} {2015})},\ \Eprint
  {http://arxiv.org/abs/1503.01469} {arXiv:1503.01469 [hep-ph]}\BibitemShut
  {NoStop}%
\bibitem [{\citenamefont {Frixione}\ \emph {et~al.}(1996)\citenamefont
  {Frixione}, \citenamefont {Kunszt},\ and\ \citenamefont
  {Signer}}]{Frixione:1995ms}%
  \BibitemOpen
  \bibfield  {author} {\bibinfo {author} {\bibfnamefont {S.}~\bibnamefont
  {Frixione}}, \bibinfo {author} {\bibfnamefont {Z.}~\bibnamefont {Kunszt}}, \
  and\ \bibinfo {author} {\bibfnamefont {A.}~\bibnamefont {Signer}},\ }\href
  {\doibase 10.1016/0550-3213(96)00110-1} {\bibfield  {journal} {\bibinfo
  {journal} {Nucl. Phys. B}\ }\textbf {\bibinfo {volume} {467}},\ \bibinfo
  {pages} {399} (\bibinfo {year} {1996})},\ \Eprint
  {http://arxiv.org/abs/hep-ph/9512328} {arXiv:hep-ph/9512328}\BibitemShut
  {NoStop}%
\bibitem [{\citenamefont {Frixione}(1997)}]{Frixione:1997np}%
  \BibitemOpen
  \bibfield  {author} {\bibinfo {author} {\bibfnamefont {S.}~\bibnamefont
  {Frixione}},\ }\href {\doibase 10.1016/S0550-3213(97)00574-9} {\bibfield
  {journal} {\bibinfo  {journal} {Nucl. Phys. B}\ }\textbf {\bibinfo {volume}
  {507}},\ \bibinfo {pages} {295} (\bibinfo {year} {1997})},\ \Eprint
  {http://arxiv.org/abs/hep-ph/9706545} {arXiv:hep-ph/9706545}\BibitemShut
  {NoStop}%
\bibitem [{\citenamefont {Crivellin}\ \emph
  {et~al.}(2021{\natexlab{d}})\citenamefont {Crivellin}, \citenamefont {Greub},
  \citenamefont {M\"uller},\ and\ \citenamefont
  {Saturnino}}]{Crivellin:2020mjs}%
  \BibitemOpen
  \bibfield  {author} {\bibinfo {author} {\bibfnamefont {A.}~\bibnamefont
  {Crivellin}}, \bibinfo {author} {\bibfnamefont {C.}~\bibnamefont {Greub}},
  \bibinfo {author} {\bibfnamefont {D.}~\bibnamefont {M\"uller}}, \ and\
  \bibinfo {author} {\bibfnamefont {F.}~\bibnamefont {Saturnino}},\ }\href
  {\doibase 10.1007/JHEP02(2021)182} {\bibfield  {journal} {\bibinfo  {journal}
  {JHEP}\ }\textbf {\bibinfo {volume} {02}},\ \bibinfo {pages} {182} (\bibinfo
  {year} {2021}{\natexlab{d}})},\ \Eprint {http://arxiv.org/abs/2010.06593}
  {arXiv:2010.06593 [hep-ph]}\BibitemShut {NoStop}%
\bibitem [{\citenamefont {Ball}\ \emph {et~al.}(2022)\citenamefont {Ball} \emph
  {et~al.}}]{Ball:2021leu}%
  \BibitemOpen
  \bibfield  {author} {\bibinfo {author} {\bibfnamefont {R.~D.}\ \bibnamefont
  {Ball}} \emph {et~al.} (\bibinfo {collaboration} {NNPDF}),\ }\href {\doibase
  10.1140/epjc/s10052-022-10328-7} {\bibfield  {journal} {\bibinfo  {journal}
  {Eur. Phys. J. C}\ }\textbf {\bibinfo {volume} {82}},\ \bibinfo {pages} {428}
  (\bibinfo {year} {2022})},\ \Eprint {http://arxiv.org/abs/2109.02653}
  {arXiv:2109.02653 [hep-ph]} \BibitemShut {NoStop}%
\bibitem [{\citenamefont {Sj\"ostrand}\ \emph {et~al.}(2015)\citenamefont
  {Sj\"ostrand}, \citenamefont {Ask}, \citenamefont {Christiansen},
  \citenamefont {Corke}, \citenamefont {Desai}, \citenamefont {Ilten},
  \citenamefont {Mrenna}, \citenamefont {Prestel}, \citenamefont {Rasmussen},\
  and\ \citenamefont {Skands}}]{Sjostrand:2014zea}%
  \BibitemOpen
  \bibfield  {author} {\bibinfo {author} {\bibfnamefont {T.}~\bibnamefont
  {Sj\"ostrand}}, \bibinfo {author} {\bibfnamefont {S.}~\bibnamefont {Ask}},
  \bibinfo {author} {\bibfnamefont {J.~R.}\ \bibnamefont {Christiansen}},
  \bibinfo {author} {\bibfnamefont {R.}~\bibnamefont {Corke}}, \bibinfo
  {author} {\bibfnamefont {N.}~\bibnamefont {Desai}}, \bibinfo {author}
  {\bibfnamefont {P.}~\bibnamefont {Ilten}}, \bibinfo {author} {\bibfnamefont
  {S.}~\bibnamefont {Mrenna}}, \bibinfo {author} {\bibfnamefont
  {S.}~\bibnamefont {Prestel}}, \bibinfo {author} {\bibfnamefont {C.~O.}\
  \bibnamefont {Rasmussen}}, \ and\ \bibinfo {author} {\bibfnamefont {P.~Z.}\
  \bibnamefont {Skands}},\ }\href {\doibase 10.1016/j.cpc.2015.01.024}
  {\bibfield  {journal} {\bibinfo  {journal} {Comput. Phys. Commun.}\ }\textbf
  {\bibinfo {volume} {191}},\ \bibinfo {pages} {159} (\bibinfo {year}
  {2015})},\ \Eprint {http://arxiv.org/abs/1410.3012} {arXiv:1410.3012
  [hep-ph]}\BibitemShut {NoStop}%
\bibitem [{\citenamefont {Ciafaloni}\ and\ \citenamefont
  {Comelli}(1999)}]{Ciafaloni:1998xg}%
  \BibitemOpen
  \bibfield  {author} {\bibinfo {author} {\bibfnamefont {P.}~\bibnamefont
  {Ciafaloni}}\ and\ \bibinfo {author} {\bibfnamefont {D.}~\bibnamefont
  {Comelli}},\ }\href {\doibase 10.1016/S0370-2693(98)01541-X} {\bibfield
  {journal} {\bibinfo  {journal} {Phys. Lett. B}\ }\textbf {\bibinfo {volume}
  {446}},\ \bibinfo {pages} {278} (\bibinfo {year} {1999})},\ \Eprint
  {http://arxiv.org/abs/hep-ph/9809321} {arXiv:hep-ph/9809321}\BibitemShut
  {NoStop}%
\bibitem [{\citenamefont {Cacciari}\ \emph {et~al.}(2008)\citenamefont
  {Cacciari}, \citenamefont {Salam},\ and\ \citenamefont
  {Soyez}}]{Cacciari:2008gp}%
  \BibitemOpen
  \bibfield  {author} {\bibinfo {author} {\bibfnamefont {M.}~\bibnamefont
  {Cacciari}}, \bibinfo {author} {\bibfnamefont {G.~P.}\ \bibnamefont {Salam}},
  \ and\ \bibinfo {author} {\bibfnamefont {G.}~\bibnamefont {Soyez}},\ }\href
  {\doibase 10.1088/1126-6708/2008/04/063} {\bibfield  {journal} {\bibinfo
  {journal} {JHEP}\ }\textbf {\bibinfo {volume} {04}},\ \bibinfo {pages} {063}
  (\bibinfo {year} {2008})},\ \Eprint {http://arxiv.org/abs/0802.1189}
  {arXiv:0802.1189 [hep-ph]}\BibitemShut {NoStop}%
\bibitem [{\citenamefont {Cacciari}\ \emph {et~al.}(2012)\citenamefont
  {Cacciari}, \citenamefont {Salam},\ and\ \citenamefont
  {Soyez}}]{Cacciari:2011ma}%
  \BibitemOpen
  \bibfield  {author} {\bibinfo {author} {\bibfnamefont {M.}~\bibnamefont
  {Cacciari}}, \bibinfo {author} {\bibfnamefont {G.~P.}\ \bibnamefont {Salam}},
  \ and\ \bibinfo {author} {\bibfnamefont {G.}~\bibnamefont {Soyez}},\ }\href
  {\doibase 10.1140/epjc/s10052-012-1896-2} {\bibfield  {journal} {\bibinfo
  {journal} {Eur. Phys. J. C}\ }\textbf {\bibinfo {volume} {72}},\ \bibinfo
  {pages} {1896} (\bibinfo {year} {2012})},\ \Eprint
  {http://arxiv.org/abs/1111.6097} {arXiv:1111.6097 [hep-ph]}\BibitemShut
  {NoStop}%
\bibitem [{\citenamefont {Aad}\ \emph {et~al.}(2019)\citenamefont {Aad} \emph
  {et~al.}}]{ATLAS:2019bwq}%
  \BibitemOpen
  \bibfield  {author} {\bibinfo {author} {\bibfnamefont {G.}~\bibnamefont
  {Aad}} \emph {et~al.} (\bibinfo {collaboration} {ATLAS}),\ }\href {\doibase
  10.1140/epjc/s10052-019-7450-8} {\bibfield  {journal} {\bibinfo  {journal}
  {Eur. Phys. J. C}\ }\textbf {\bibinfo {volume} {79}},\ \bibinfo {pages} {970}
  (\bibinfo {year} {2019})},\ \Eprint {http://arxiv.org/abs/1907.05120}
  {arXiv:1907.05120 [hep-ex]}\BibitemShut {NoStop}%
\bibitem [{\citenamefont {Aad}\ \emph {et~al.}(2016)\citenamefont {Aad} \emph
  {et~al.}}]{ATLAS:2016lqx}%
  \BibitemOpen
  \bibfield  {author} {\bibinfo {author} {\bibfnamefont {G.}~\bibnamefont
  {Aad}} \emph {et~al.} (\bibinfo {collaboration} {ATLAS}),\ }\href {\doibase
  10.1140/epjc/s10052-016-4120-y} {\bibfield  {journal} {\bibinfo  {journal}
  {Eur. Phys. J. C}\ }\textbf {\bibinfo {volume} {76}},\ \bibinfo {pages} {292}
  (\bibinfo {year} {2016})},\ \Eprint {http://arxiv.org/abs/1603.05598}
  {arXiv:1603.05598 [hep-ex]}\BibitemShut {NoStop}%
\bibitem [{\citenamefont {Aad}\ \emph {et~al.}(2020{\natexlab{c}})\citenamefont
  {Aad} \emph {et~al.}}]{ATLAS:2020gty}%
  \BibitemOpen
  \bibfield  {author} {\bibinfo {author} {\bibfnamefont {G.}~\bibnamefont
  {Aad}} \emph {et~al.} (\bibinfo {collaboration} {ATLAS}),\ }\href {\doibase
  10.1088/1748-0221/15/09/p09015} {\bibfield  {journal} {\bibinfo  {journal}
  {JINST}\ }\textbf {\bibinfo {volume} {15}},\ \bibinfo {pages} {P09015}
  (\bibinfo {year} {2020}{\natexlab{c}})},\ \Eprint
  {http://arxiv.org/abs/2004.13447} {arXiv:2004.13447 [physics.ins-det]}
 \BibitemShut {NoStop}%
\bibitem [{\citenamefont {Conte}\ \emph {et~al.}(2013)\citenamefont {Conte},
  \citenamefont {Fuks},\ and\ \citenamefont {Serret}}]{Conte:2012fm}%
  \BibitemOpen
  \bibfield  {author} {\bibinfo {author} {\bibfnamefont {E.}~\bibnamefont
  {Conte}}, \bibinfo {author} {\bibfnamefont {B.}~\bibnamefont {Fuks}}, \ and\
  \bibinfo {author} {\bibfnamefont {G.}~\bibnamefont {Serret}},\ }\href
  {\doibase 10.1016/j.cpc.2012.09.009} {\bibfield  {journal} {\bibinfo
  {journal} {Comput. Phys. Commun.}\ }\textbf {\bibinfo {volume} {184}},\
  \bibinfo {pages} {222} (\bibinfo {year} {2013})},\ \Eprint
  {http://arxiv.org/abs/1206.1599} {arXiv:1206.1599 [hep-ph]}\BibitemShut
  {NoStop}%
\bibitem [{\citenamefont {de~Favereau}\ \emph {et~al.}(2014)\citenamefont
  {de~Favereau}, \citenamefont {Delaere}, \citenamefont {Demin}, \citenamefont
  {Giammanco}, \citenamefont {Lema\^\i{}tre}, \citenamefont {Mertens},\ and\
  \citenamefont {Selvaggi}}]{deFavereau:2013fsa}%
  \BibitemOpen
  \bibfield  {author} {\bibinfo {author} {\bibfnamefont {J.}~\bibnamefont
  {de~Favereau}}, \bibinfo {author} {\bibfnamefont {C.}~\bibnamefont
  {Delaere}}, \bibinfo {author} {\bibfnamefont {P.}~\bibnamefont {Demin}},
  \bibinfo {author} {\bibfnamefont {A.}~\bibnamefont {Giammanco}}, \bibinfo
  {author} {\bibfnamefont {V.}~\bibnamefont {Lema\^\i{}tre}}, \bibinfo {author}
  {\bibfnamefont {A.}~\bibnamefont {Mertens}}, \ and\ \bibinfo {author}
  {\bibfnamefont {M.}~\bibnamefont {Selvaggi}} (\bibinfo {collaboration}
  {DELPHES 3}),\ }\href {\doibase 10.1007/JHEP02(2014)057} {\bibfield
  {journal} {\bibinfo  {journal} {JHEP}\ }\textbf {\bibinfo {volume} {02}},\
  \bibinfo {pages} {057} (\bibinfo {year} {2014})},\ \Eprint
  {http://arxiv.org/abs/1307.6346} {arXiv:1307.6346 [hep-ex]}\BibitemShut
  {NoStop}%
\bibitem [{\citenamefont {Tumasyan}\ \emph
  {et~al.}(2022{\natexlab{b}})\citenamefont {Tumasyan} \emph
  {et~al.}}]{CMS:2022prd}%
  \BibitemOpen
  \bibfield  {author} {\bibinfo {author} {\bibfnamefont {A.}~\bibnamefont
  {Tumasyan}} \emph {et~al.} (\bibinfo {collaboration} {CMS}),\ }\href@noop {}
  {\  (\bibinfo {year} {2022}{\natexlab{b}})},\ \Eprint
  {http://arxiv.org/abs/2201.08458} {arXiv:2201.08458 [hep-ex]}\BibitemShut
  {NoStop}%
\bibitem [{\citenamefont {Sirunyan}\ \emph {et~al.}(2018)\citenamefont
  {Sirunyan} \emph {et~al.}}]{CMS:2017wtu}%
  \BibitemOpen
  \bibfield  {author} {\bibinfo {author} {\bibfnamefont {A.~M.}\ \bibnamefont
  {Sirunyan}} \emph {et~al.} (\bibinfo {collaboration} {CMS}),\ }\href
  {\doibase 10.1088/1748-0221/13/05/P05011} {\bibfield  {journal} {\bibinfo
  {journal} {JINST}\ }\textbf {\bibinfo {volume} {13}},\ \bibinfo {pages}
  {P05011} (\bibinfo {year} {2018})},\ \Eprint
  {http://arxiv.org/abs/1712.07158} {arXiv:1712.07158 [physics.ins-det]}
 \BibitemShut {NoStop}%
\bibitem [{\citenamefont {Bols}\ \emph {et~al.}(2020)\citenamefont {Bols},
  \citenamefont {Kieseler}, \citenamefont {Verzetti}, \citenamefont {Stoye},\
  and\ \citenamefont {Stakia}}]{Bols:2020bkb}%
  \BibitemOpen
  \bibfield  {author} {\bibinfo {author} {\bibfnamefont {E.}~\bibnamefont
  {Bols}}, \bibinfo {author} {\bibfnamefont {J.}~\bibnamefont {Kieseler}},
  \bibinfo {author} {\bibfnamefont {M.}~\bibnamefont {Verzetti}}, \bibinfo
  {author} {\bibfnamefont {M.}~\bibnamefont {Stoye}}, \ and\ \bibinfo {author}
  {\bibfnamefont {A.}~\bibnamefont {Stakia}},\ }\href {\doibase
  10.1088/1748-0221/15/12/P12012} {\bibfield  {journal} {\bibinfo  {journal}
  {JINST}\ }\textbf {\bibinfo {volume} {15}},\ \bibinfo {pages} {P12012}
  (\bibinfo {year} {2020})},\ \Eprint {http://arxiv.org/abs/2008.10519}
  {arXiv:2008.10519 [hep-ex]}\BibitemShut {NoStop}%
\bibitem [{\citenamefont {Aad}\ \emph {et~al.}(2014)\citenamefont {Aad} \emph
  {et~al.}}]{ATLAS:2014vhc}%
  \BibitemOpen
  \bibfield  {author} {\bibinfo {author} {\bibfnamefont {G.}~\bibnamefont
  {Aad}} \emph {et~al.} (\bibinfo {collaboration} {ATLAS}),\ }\href {\doibase
  10.1007/JHEP11(2014)056} {\bibfield  {journal} {\bibinfo  {journal} {JHEP}\
  }\textbf {\bibinfo {volume} {11}},\ \bibinfo {pages} {056} (\bibinfo {year}
  {2014})},\ \Eprint {http://arxiv.org/abs/1409.6064} {arXiv:1409.6064
  [hep-ex]}\BibitemShut {NoStop}%
\bibitem [{\citenamefont {Aad}\ \emph {et~al.}(2020{\natexlab{d}})\citenamefont
  {Aad} \emph {et~al.}}]{ATLAS:2020dsk}%
  \BibitemOpen
  \bibfield  {author} {\bibinfo {author} {\bibfnamefont {G.}~\bibnamefont
  {Aad}} \emph {et~al.} (\bibinfo {collaboration} {ATLAS}),\ }\href {\doibase
  10.1007/JHEP10(2020)112} {\bibfield  {journal} {\bibinfo  {journal} {JHEP}\
  }\textbf {\bibinfo {volume} {10}},\ \bibinfo {pages} {112} (\bibinfo {year}
  {2020}{\natexlab{d}})},\ \Eprint {http://arxiv.org/abs/2006.05872}
  {arXiv:2006.05872 [hep-ex]}\BibitemShut {NoStop}%
\bibitem [{\citenamefont {Aad}\ \emph {et~al.}(2021{\natexlab{b}})\citenamefont
  {Aad} \emph {et~al.}}]{ATLAS:2021jyv}%
  \BibitemOpen
  \bibfield  {author} {\bibinfo {author} {\bibfnamefont {G.}~\bibnamefont
  {Aad}} \emph {et~al.} (\bibinfo {collaboration} {ATLAS}),\ }\href {\doibase
  10.1103/PhysRevD.104.112005} {\bibfield  {journal} {\bibinfo  {journal}
  {Phys. Rev. D}\ }\textbf {\bibinfo {volume} {104}},\ \bibinfo {pages}
  {112005} (\bibinfo {year} {2021}{\natexlab{b}})},\ \Eprint
  {http://arxiv.org/abs/2108.07665} {arXiv:2108.07665 [hep-ex]}\BibitemShut
  {NoStop}%
\bibitem [{\citenamefont {Cowan}\ \emph {et~al.}(2011)\citenamefont {Cowan},
  \citenamefont {Cranmer}, \citenamefont {Gross},\ and\ \citenamefont
  {Vitells}}]{Cowan:2010js}%
  \BibitemOpen
  \bibfield  {author} {\bibinfo {author} {\bibfnamefont {G.}~\bibnamefont
  {Cowan}}, \bibinfo {author} {\bibfnamefont {K.}~\bibnamefont {Cranmer}},
  \bibinfo {author} {\bibfnamefont {E.}~\bibnamefont {Gross}}, \ and\ \bibinfo
  {author} {\bibfnamefont {O.}~\bibnamefont {Vitells}},\ }\href {\doibase
  10.1140/epjc/s10052-011-1554-0} {\bibfield  {journal} {\bibinfo  {journal}
  {Eur. Phys. J. C}\ }\textbf {\bibinfo {volume} {71}},\ \bibinfo {pages}
  {1554} (\bibinfo {year} {2011})},\ \bibinfo {note} {[Erratum: Eur. Phys. J. C
  {\bf 73}, 2501 (2013)]},\ \Eprint {http://arxiv.org/abs/1007.1727} {arXiv:1007.1727
  [physics.data-an]}\BibitemShut {NoStop}%
\bibitem [{POW()}]{POWHEGBOX}%
  \BibitemOpen
  \href {http://powhegbox.mib.infn.it} {\emph {\bibinfo {title} {{The POWHEG
  BOX}}}}.\BibitemShut {Stop}%
\bibitem [{\citenamefont {Collins}\ and\ \citenamefont
  {Soper}(1977)}]{Collins:1977iv}%
  \BibitemOpen
  \bibfield  {author} {\bibinfo {author} {\bibfnamefont {J.~C.}\ \bibnamefont
  {Collins}}\ and\ \bibinfo {author} {\bibfnamefont {D.~E.}\ \bibnamefont
  {Soper}},\ }\href {\doibase 10.1103/PhysRevD.16.2219} {\bibfield  {journal}
  {\bibinfo  {journal} {Phys. Rev. D}\ }\textbf {\bibinfo {volume} {16}},\
  \bibinfo {pages} {2219} (\bibinfo {year} {1977})}\BibitemShut {NoStop}%
\bibitem [{\citenamefont {Binosi}\ \emph {et~al.}(2009)\citenamefont {Binosi},
  \citenamefont {Collins}, \citenamefont {Kaufhold},\ and\ \citenamefont
  {Theussl}}]{Binosi:2008ig}%
  \BibitemOpen
  \bibfield  {author} {\bibinfo {author} {\bibfnamefont {D.}~\bibnamefont
  {Binosi}}, \bibinfo {author} {\bibfnamefont {J.}~\bibnamefont {Collins}},
  \bibinfo {author} {\bibfnamefont {C.}~\bibnamefont {Kaufhold}}, \ and\
  \bibinfo {author} {\bibfnamefont {L.}~\bibnamefont {Theussl}},\ }\href
  {\doibase 10.1016/j.cpc.2009.02.020} {\bibfield  {journal} {\bibinfo
  {journal} {Comput. Phys. Commun.}\ }\textbf {\bibinfo {volume} {180}},\
  \bibinfo {pages} {1709} (\bibinfo {year} {2009})},\ \Eprint
  {http://arxiv.org/abs/0811.4113} {arXiv:0811.4113 [hep-ph]}\BibitemShut
  {NoStop}%
\bibitem [{\citenamefont {Qu}\ and\ \citenamefont
  {Gouskos}(2020)}]{Qu:2019gqs}%
  \BibitemOpen
  \bibfield  {author} {\bibinfo {author} {\bibfnamefont {H.}~\bibnamefont
  {Qu}}\ and\ \bibinfo {author} {\bibfnamefont {L.}~\bibnamefont {Gouskos}},\
  }\href {\doibase 10.1103/PhysRevD.101.056019} {\bibfield  {journal} {\bibinfo
   {journal} {Phys. Rev. D}\ }\textbf {\bibinfo {volume} {101}},\ \bibinfo
  {pages} {056019} (\bibinfo {year} {2020})},\ \Eprint
  {http://arxiv.org/abs/1902.08570} {arXiv:1902.08570 [hep-ph]}\BibitemShut
  {NoStop}%
\bibitem [{\citenamefont {Aad}\ \emph {et~al.}(2022)\citenamefont {Aad} \emph
  {et~al.}}]{ATLAS:2022ers}%
  \BibitemOpen
  \bibfield  {author} {\bibinfo {author} {\bibfnamefont {G.}~\bibnamefont
  {Aad}} \emph {et~al.} (\bibinfo {collaboration} {ATLAS}),\ }\href@noop {} {\
  (\bibinfo {year} {2022})},\ \Eprint {http://arxiv.org/abs/2201.11428}
  {arXiv:2201.11428 [hep-ex]}\BibitemShut {NoStop}%
\bibitem [{\citenamefont {Tumasyan}\ \emph
  {et~al.}(2022{\natexlab{c}})\citenamefont {Tumasyan} \emph
  {et~al.}}]{CMS:2022psv}%
  \BibitemOpen
  \bibfield  {author} {\bibinfo {author} {\bibfnamefont {A.}~\bibnamefont
  {Tumasyan}} \emph {et~al.} (\bibinfo {collaboration} {CMS}),\ }\href@noop {}
  {\  (\bibinfo {year} {2022}{\natexlab{c}})},\ \Eprint
  {http://arxiv.org/abs/2205.05550} {arXiv:2205.05550 [hep-ex]}\BibitemShut
  {NoStop}%
\end{thebibliography}

%

\end{document}